\documentclass[reqno]{amsart}
\usepackage{hyperref}
\usepackage{amsmath,amssymb} 

\usepackage{euscript}
\usepackage[T2A]{fontenc}
\usepackage[utf8]{inputenc}
\usepackage{textcomp}

\allowdisplaybreaks

\theoremstyle{plain}
\newtheorem{theorem}{Theorem}[section]

\theoremstyle{definition}

\theoremstyle{remark}
\newtheorem{remark}[theorem]{Remark}
\numberwithin{equation}{section}

\begin{document}
	
	\title[ Disorder solutions for the free energy of the  Ising-like models ]
	{Disorder solutions for the free energy of the  Ising-like models	}
	
	\author[P. V. Khrapov]{Pavel V. Khrapov}
	\address{Pavel V. Khrapov, Department of Mathematics, Bauman Moscow State Technical University (5/1 2-nd
Baumanskaya St., Moscow 105005, Russia)
}  
	\email{khrapov@bmstu.ru }	



	\begin{abstract}
		For arbitrary Ising-like models of any dimension and Hamiltonians with a
		finite support with all possible multispin interactions and boundary conditions with a shift, 
		the exact value of the free energy in the thermodynamic limit is obtained at some parametrically 
		specified set of multispin interaction coefficients. In this case, half of the multispin interaction 
		coefficients and the coordinates of the special eigenvector corresponding to the largest eigenvalue 
		of the elementary transfer matrix are parameters, and the second half of the multispin coefficients 
		is calculated using simple explicit formulas. For models with Hamiltonians invariant under the reversal 
		of signs of all spins, the formulas are simplified. As examples of independent interest, solutions 
		are written for the cases when the support of the Hamiltonian is a simplex, a cube, the support 
		of the ANNNI model in spaces of 2, 3 and arbitrary dimensions.
\end{abstract}
	\maketitle

		\section{Introduction}
		\label{introduction}
		 
	Ising model with interactions between pairs of nearest neighbours is one of the most studied systems in statistical mechanics. Some exact solutions ( with an analitical formulae for the partition function or free energy of the system) were obtained mainly for planar models, among these the exact Onsager's solution \cite{Onsager} of two-dimensional Ising model without an external magnetic field stands out clearly.
		At the present time there are a lot of different anisotropic models like Ising or Potts models with different coupling constants in the different directions, for which the wonderful subsets   called "disorder solutions" are found in the space of parameters, where the partition function can be calculated and represented in the simple form .  Remarkable examples were provided in the case of anisotropic models. Stephenson J. \cite{Stephenson} explicitly researched pair correlations between spins at the sites of the anisotropic triangular lattice along the axes, Enting I.G. \cite{Enting}  showed that the ratios of certain triplet order parameters to magnetisation in
		honeycomb and diamond lattice Ising models can be easily calculated  , 
		Baxter R.J. \cite{Baxter_1984} analyzed the disorder varieties of 
		the Ising model with all possible interactions around a face of
		the square lattice,    "interactions-round-a-face" (IRF) model
		on the square lattice. For the disorder solutions expressions obtained for the free energy and intra-row correlations. These are applied to the checkerboard Potts model. Ruj$\acute a$n P. \cite{RujAn_1}, \cite{RujAn_2}, \cite{RujAn_3} researched the 
		IRF model on the square lattice. He examined the general eight-vertex model thoroughly. Wu F.Y. \cite{Wu1985} found  the disorder solutions for the "checkerboard-triangular" lattice. 
		M. T. Jaekel and J. M. Maillard \cite{Jaekel_Maillard_1} found a local criterion which characterizes  disorder varieties for any dimensionality and explains the effective dimensional reduction occurring in the model. They found disorder solutions for several models,
	 	such as anisotropic triangular Ising model with the field, checkerboard Potts model, Potts model on the Kagome lattice and general anisotropic cubic Ising model with nearest neighbour interactions, using a  star-triangle transformation.		
			Dhar D., Maillard J.M. \cite{Dhar_Maillard} ,   
		Georges A., Hansel D., Doussal P. L., Maillard J. M. \cite{Georges_Hansel_Doussal_Maillard_1987} used this local criterion to calculate
		correlation functions on the disorder varieties of Ising and Potts models.
		Meyer H., Angl\`{e}s  d’Auriac J.-C.,  Maillard J.M.  \cite{Meyer_et_al_1997} studied the disorder varieties
		of the eight vertex model in the framework of a random
		matrix theory approach to the transfer matrix.
		Various methods were used to obtain these solutions:
		methods related to crystal growth (Enting I.G.  \cite{Enting},  Welberry T.R., Galbraith R. \cite{Welberry_Galbraith}, Welberry T.R., Miller G.H. \cite{Welberry_Miller}), to Markov
		processes (Verhagen A.M.W. \cite{Verhagen}) and 
		transfer matrix technique ( Ruj$\acute a$n P. \cite{RujAn_2}, \cite{RujAn_3}, Baxter, R.J. \cite{Baxter_1984}, Minlos R.A., Sinai Y.G.\cite{Minlos_Sinai},
		Malyshev, V.A., Minlos, R.A. \cite{Malyshev_Minlos},  Minlos R.A., Khrapov P.V. \cite{Minlos_Khrapov}). In the most cases, the problem of calculating the partition function was compared with an equivalent one in another area, where appropriate methods were available to find a solution.
		Reviews on this theme can be found in Wu F.Y.  \cite{Wu2009exactly} , Baxter R.J.  \cite{Baxter2016}, Pelizzola A. \cite{Pelizzola_2000}, \cite{Pelizzola_2005}.
		This article is a logical continuation of the author’s works \cite{Khrapov2}, \cite{Khrapov3}, \cite{Khrapov4},  \cite{Khrapov5},  \cite{Khrapov6},  \cite{Khrapov7} which outlines a general methodology for finding such disordered solutions for generalized Ising and Potts models, and explicitly some of these solutions for a square and three-dimensional generalized Ising model are obtained. 
		In this work, for Ising-like models 
		 of any dimension and with the Hamiltonian with finite support with all possible multispin interactions the explicit exact formulas for finding partition function and free
		    energy in the thermodynamic limit were derived, in a parametric dependence on half of the multispin interaction coefficients of the Hamiltonian and the coordinates of a special eigenvector corresponding to the largest eigenvalue of the elementary transfer matrix. The formulas are simplified for the models with the  Hamiltonians invariant under the reversal of signs of all spins. By passing to the limit, the same results are carried over to Ising-like models, for which the supports are unlimited. Without loss of generality, the use of the formulas is demonstrated on specific examples that are of independent interest, and have been repeatedly described in the scientific literature. For the partition function and  free energy in the thermodynamic limit disorder solutions are found for Ising-like models with the support of the Hamiltonian in the form of $\nu$-dimensional simplex, in the form of $\nu$ - dimensional cube and for generalized $\nu$ - dimensional model ANNNI. Formulas are separately adapted for $\nu=2,3 $  for models with the support of the Hamiltonian in the form of a simplex and a cube. For 2D simplex, or "checkerboard-triangular" Ising model, solutions is compared with solutions from   Wu F.Y.  \cite{Wu1985}, their coincidence is obtained on numerical examples. Along the way, this shows that in the thermodynamic limit the model from \cite{Wu1985} coincides with the models in this paper. For $\nu=2 $
		   in this work the IRF model on the square lattice is given in "interaction representation" rather than in the "weight (Boltzmann) representation"
		   in the articles \cite{RujAn_2}, \cite{RujAn_3},  \cite{Baxter_1984}.  The disorder solutions were found on the 8-dimensional subset of 10-dimensional space of all independent parameters. This is the generalization of results from \cite{Khrapov6}, where the solutions were found on the 7-dimensional subset of the 10-dimensional space of all independent parameters, and is the same as the result \cite{Khrapov7}, but in this article another parameterization is given, which seems to be more convenient and universal. Numerical examples are given to make it easier for the reader to check the correctness of the exact analytical solutions found.	
		 
		%%%%%%%%%%%%%%%%%%%%%%%%%%%%%%%%%
		
		This work has the following stucture.		
		
		 The section \ref{Model_description} "Model description and main results" is devoted to the description of the model, the formation of an elementary transfer matrix, a system of equations for the largest eigenvalue of the transfer matrix, reduction of the system of equations to a system of linear  equations, the solution of this system of equations and obtaining the calculated coefficients of multi-spin interaction. 
		 In subsection \ref{Model_description_lattice} "Model description" the  $\nu$ - dimensional lattice model with helical boundary conditions were described.
		Toroidal boundary conditions \cite{Khrapov3} are with a shift by one (similar to helical ones), and a cyclic closure of the set of all points (in natural ordering).
		For these model the Hamiltonian were written and converted into a more suitable form for further research.
		In  subsection  \ref{Partition_function_arbitrary} "Free energy for models with a Hamiltonian on an arbitrary bounded support"
		sparse elementary transfer matrices  \cite{Khrapov3} with non-negative elements were constructed. For arbitrary the Hamiltonian with bounded support the system of equations for the largest eigenvalue of the elementary transfer matrix $\lambda_{max}$ is written, reduced to a system of linear  equations and solved in general form for eigenvectors of a special form. Half of the multi-spin interaction coefficients and the coordinates of the elementary transfer matrix eigenvector corresponding to the largest eigenvalue are used as the solution parameters.
	  	Subsection \ref{even} "Free energy for models with the Hamiltonians invariant under the reversal of signs of all spins" is devoted to finding disorder solutions for the partition function and free energy of models with the Hamiltonians, invariant under the reversal of signs of all spins. In our case, the Hamiltonian of such models has the form (\ref{Hamiltonian1}), where  $s=2 p $ are only even.
	  	The general system of equations is simplified and solved, taking into account the specifics of the Hamiltonian.
	  	Then some examples of independent interest are considered.
				 
		 In section \ref {nuD_simplex} "Free energy  for $\nu$ - dimensional models with support of the Hamiltonian in the form of  $\nu$ - dimensional simplex" the general theory from \ref {Partition_function_arbitrary} and \ref {even} is applied to models in which the support of the Hamiltonian has the form of a $ \nu $ - dimensional simplex.
		  The general parametric solution  has been adapted for the Hamiltonians with an arbitrary multispin interaction (\ref{K_nuD_simplex}) and for  the Hamiltonians invariant under the reversal of signs of all spins  (\ref{K_nuD_simplex_even}).	
		 In subsection  
		\ref{checkerboard-triangular}  "Free energy of "checkerboard-triangular" model on the triangular lattice" the general Hamiltonian (\ref{Chess_Hamiltonian1}) is written for  "checkerboard-triangular" model on the triangular lattice. In subsubsection    \ref{checkerboard-triangular-main}  the formulas (\ref{Chess_K}) to find three interaction parameters of the general Hamiltonian (\ref{Chess_Hamiltonian1}) are written in a parametric dependence on the four remaining parameters of the Hamiltonian, taking any real values, and   coordinates of the eigenvector of the elementary transfer matrix corresponding to the largest eigenvalue. A specific example with numeric parameters is considered. It is shown that this numerical solution matches the solution from \cite {Wu1985}, in which disordered solutions are obtained for "checkerboard-triangular" model on the triangular lattice with periodic boundary conditions.   Along the way, this indirectly shows that in the thermodynamic limit the model from \cite {Wu1985} coincides with the "checkerboard-triangular" model in this paper.
		In remark \ref{remark_chess} it is shown that in this numerical example, when the components of the eigenvector of the elementary transfer matrix change, the total external field and the rest of the calculated  parameters of the Hamiltonian (\ref{Chess_Hamiltonian1}) will not change.
		In  subsubsection \ref{checkerboard-triangular-paired} formulas were written for finding the interaction parameter of the Hamiltonian  (\ref{Chess_Hamiltonian_paired}) containing only pair interactions,  in a parametric dependence on the two remaining parameters of the Hamiltonian, taking any real values, and the  components of the eigenvector of the elementary transfer matrix corresponding to the largest eigenvalue of the transfer matrix.		
		In subsection \ref {3D_simplex}, general formulas from section \ref {nuD_simplex} are given for three-dimensional simplex models of the Ising  type with support in the form of a three-dimensional simplex for Hamiltonians of both the most general form and those invariant under the change of sign of all spins.
		 The system of 8 equations is written for finding the dependent parameters of the Hamiltonian and its solution.  A separate solution has also been written for finding the dependent parameters of the Hamiltonian for models with a Hamiltonian supported by a three-dimensional simplex and invariant under the change of sign of all spins.

		 In the section \ref {nuD} "Free energy for $ \nu $ - dimensional models with the support of the Hamiltonian in the form of $ \nu $ - dimensional cube", the general theory from (\ref {Partition_function_arbitrary}) and (\ref {even}) is applied to models with the support of the Hamiltonian in the form of a $ \nu $ - dimensional cube. The general parametric solution was adapted both for Hamiltonians with an arbitrary multispin interaction (\ref {K_nuD}) and for Hamiltonians that are invariant under the sign of all spins (\ref {K_even_cube}).
		 In subsection  
		 (\ref{2D})  "Free energy  for cube  model on the 2D lattice. General case" the general Hamiltonian is written for flat Ising model and all parametric solutions are considered for $\nu=2$. An example of a numerical calculation for the Hamiltonian of the most general form is given. In remark \ref{remark_2D} it is said that when the four parameters of the multispin interaction are zeroed, we get the disorder solutions on a complete triangular lattice. And when five parameters are zeroed, we get solutions on a "checkerboard-triangular" lattice.
		   In subsection 
		   \ref{2D_even}  "Free energy  for cube models invariant under the reversal of signs of all spins on the 2D lattice" the Hamiltonian invariant under the reversal of signs of all spins is written and all parametric solutions are considered for $\nu=2$.  In remark (\ref{remark_2D_even}) it is said that when two parameters are zeroed, we obtain solutions on a triangular lattice for models with the Hamiltonian invariant under the reversal of signs of all spins.		  
		  In subsection   
		 (\ref{3D})  "Free energy  for cube model on the 3D lattice. General case" the general Hamiltonian is written for 3D Ising-like model and all parametric solutions are considered for $\nu=3$.  The example \ref{3D lattice} of numerical calculation for the Hamiltonian of the most general form is given. 
		  		  In subsection  
		 (\ref{3D_cube_even})  "Free energy  for cube  models, invariant under the reversal of signs of all spins, on the 3D lattice" the Hamiltonian invariant under the reversal of signs of all spins is written and all parametric solutions are considered for $\nu=3$. The example \ref{example_3D_even_1} of a numerical calculation is given. 
		
		  In section \ref{nuD_ANNNI} "Free energy  for $\nu$ - dimensional generalized models ANNNI" general theory from (\ref{Partition_function_arbitrary}) and (\ref{even}) was applied to models, whose support of the Hamiltonian coincides with the support	$\nu$ - dimensional model ANNNI. The general parametric solution has been adapted for Hamiltonians with an arbitrary multispin interaction (\ref{K_ANNNI_nuD}) and for the Hamiltonian invariant under the reversal of signs of all spins (\ref{K_even_ANNNI}).

		 %%%%%%%%%%%%%%%%%%%%%%%%%%%%%%%%%
			
		\section{Model description and main results} 
		\label{Model_description}
		
		    In the first subsection of this section, a description of the $\nu$-dimensional lattice with helical boundary conditions and the general form of the considered Hamiltonian are given.
		    In the second subsection, the elementary transfer matrix is introduced, through the logarithm of the largest eigenvalue of which $ \lambda_{max} $ is calculated the free energy of the model in the thermodynamic limit.
		    		    
		    For an arbitrary Hamiltonian with bounded support, a system of equations is written for
		    $ \lambda_ {max} $, and the multi-spin interaction coefficients are calculated. The system of equations is reduced to a system of linear equations and is solved. Half of the multi-spin interaction coefficients and the components of the eigenvector of the elementary transfer matrix are used as the solution parameters.
		    In the third subsection, parametric solutions for the free energy are obtained for models with Hamiltonians invariants under the sign of all spins.
		\subsection{Model description} 
	\label{Model_description_lattice}
		
		Let us consider $ \nu$-dimensional lattice (more detailed description of this lattice can be found in \cite{Khrapov3})

		\begin{equation}\label{lattice_1}
			\mathcal {L}_{\nu}=\{t=(t_1,t_2,...,t_i,t_{i+1},...,t_{\nu}),t_i=0,1,...,L_i, i=1,2,...,\nu \},			
		\end{equation}
	 moreover
		
			\begin{equation}\label{lattice_2}
			\begin{gathered}
			(t_1, t_2,...,L_i,t_{i+1},...,t_{\nu}) \equiv 
			(t_1, t_2,...,0,t_{i+1}+1,...,t_{\nu}), i=1,2,..., {\nu}-1  \\
				(t_1, t_2,...,t_{\nu-1},L_{\nu})\equiv	(t_1, t_2,...,t_{\nu-1},0), \\
			(L_1,L_2-1,...,L_i - 1,...,L_{\nu} - 1)\equiv
			 (0,L_2,...,L_i - 1,...,L_{\nu} - 1)\equiv  \\
            (0,0,...,L_{\nu})\equiv (0,0,...,0) .
			\end{gathered} 
		\end{equation}
	Due to this procedure for identifying points, the lattice
	$\mathcal {L}_{\nu} $  has the size 
	 $L_1 \times L_2 \times ... \times L_{\nu} $, total number of lattice points - 	 
	 $L=L_1  L_2  ...  L_{\nu} $.

	Thus, on  $\mathcal {L}_{\nu}$ special boundary cyclic helical (with shift) conditions are imposed. Let us renumber all the points $ \mathcal {L}_{\nu} : $

			\begin{equation}\label{lattice_3}
			\begin{gathered}
			\tau^0=(0,0,...,0) ,
			\tau^1=(1,0,...,0) , \tau^2 =(2,0,...,0) ,...,\\
			\tau^{L_1}= (L_1,0,...,0)\equiv (0,1,0,...,0)
			,...,\tau^{L}=(0,0,...,0)\equiv \tau^0 .
			\end{gathered} 
		\end{equation}
		This numbering determines the natural positive direction of traversing all points and the ordering of the lattice nodes:  $ \tau^i < \tau^j  $ for $i < j$. 
 
	We will assume that at each site 
	$ t = (t_1, t_2, ..., t _ {\nu}) \in 
	 \mathcal {L}_{\nu} $ there is a particle.
	
	The state of particle is defined by spin $ \sigma_t \in X=\{+1,-1\}.$
	Denote by   $	\Omega$,  
	 \begin{equation}\label{Omega}
		\begin{gathered}
			\Omega=\{\{{\omega}^0, \omega^1,...,\omega^r \} \subset  \mathcal {L}_{\nu} , \omega^i <\omega^j  , i < j \} 
		\end{gathered} 
	 \end{equation} 
 	some fixed finite subset of points $\mathcal {L}_{\nu} $, we call it the support of the Hamiltonian,  
	the lowest point of which is  $\omega^{\min}=(0,0,...,0)=\omega^0$, and the largest one is  $\omega^{\max}=(\omega_1^{\max},\omega_2^{\max},...,\omega_{\nu}^{\max})=\omega^r$,
	 $r=|\Omega|-1$. Let us define 
	  \begin{equation}\label{Omega_tau}
	 	\begin{gathered}
	 			\Omega_{\tau^i}= \Omega+\tau^i , \;\;\;\;   
	 		\Omega'_{\tau^i}=( \Omega \setminus \omega^{\max}) +\tau^i ,
	 	\end{gathered} 
	 \end{equation}
	 
	  and 
	 \begin{equation}\label{Phi}
	 	\begin{gathered}
	 		\Phi=\Phi(\Omega)=\{{\omega}^{l_k} \in \Omega , k=1,...,m | \;\; {\omega}^{l_k}+(1,0,...,0) \in \Omega  \}=\{ \phi^1,\phi^2,...,\phi^m\},\\
	 		\Phi_{\tau^i}= 	\Phi+\tau^i .\\
	 	\end{gathered} 
	 \end{equation}
  
		The Hamiltonian of the model has the form
		\begin{equation}\label{Hamiltonian1}
			\begin{gathered}
				\mathcal {H} (\sigma )= -\sum_{i=0}^{L-1} 
			\sum_{\{\omega^{j_1},\omega^{j_2},...,\omega^{j_s}\}\subset \Omega_{\tau^i}}
			J_{\omega^{j_1} , \omega^{j_2} , ..., \omega^{j_s}}\sigma_{\omega^{j_1}}\sigma_{\omega^{j_2}}...\sigma_{\omega^{j_s}}	
				\end{gathered} 
		\end{equation}
		
		where $\tau^i=(\tau_1^i,...,\tau_{\nu}^i) \in  \mathcal {L}_{\nu}   , $ 
	  $ J_{\omega^{j_1} , \omega^{j_2} , ..., \omega^{j_s}} $
		are corresponding translation invariant coefficients of multi-spin interaction.
		Summation in (\ref{Hamiltonian1}) is possible over all possible non-empty subsets
		$\{ \omega^{j_1},\omega^{j_2},...,\omega^{j_s}\}\subset \Omega_{\tau^i} $, 
		${ \omega^{j_1} < \omega^{j_2} <... <\omega^{j_s}}. $	  This notation allows the formula   (\ref{Hamiltonian1}) to describe an arbitrary Hamiltonian with finite support. Let us notice that the subset ${\{\omega^{j_1},\omega^{j_2},...,\omega^{j_s}\}\subset \Omega_{\tau^i}}$ can occur with different $\tau^i$. In this case the Hamiltonian (\ref{Hamiltonian1}) can be written in the form  
	 
			\begin{equation}\label{Hamiltonian2}
			\begin{gathered}
				\mathcal {H} (\sigma )= -\sum_{i=0}^{L-1} 
				\sum_{\{\omega^{j_1},\omega^{j_2},...,\omega^{j_s}\}\subset \Omega_{\tau^i}}
				J'_{\omega^{j_1} , \omega^{j_2} , ..., \omega^{j_s}}\sigma_{\omega^{j_1}}\sigma_{\omega^{j_2}}...\sigma_{\omega^{j_s}}	
			\end{gathered} 
		\end{equation}
		where
		
			\begin{equation}\label{Hamiltonian3}
			\begin{gathered}
				J'_{\omega^{j_1} , \omega^{j_2} , ..., \omega^{j_s}}=	\sum_{\tau^i : \{\omega^{j_1},\omega^{j_2},...,\omega^{j_s}\}\subset \Omega_{\tau^i}} 
					J_{\omega^{j_1} , \omega^{j_2} , ..., \omega^{j_s}}				
			\end{gathered} 
		\end{equation}
	and summation in (\ref{Hamiltonian2}) is carried out only over different (not coinciding in case of parallel transfer) subsets
	 $\{\omega^{j_1} , \omega^{j_2} , ..., \omega^{j_s}\} \in \Omega_{\tau^i} $: of all subsets that coincide in parallel transfer, we will leave only, for example, the first in the natural  ordering. We will further consider the Hamiltonian  (\ref{Hamiltonian1}).
		
		Let us define the coefficients 
			\begin{equation}\label{Hamiltonian4}
			\begin{gathered}
				K_{\omega^{j_1} , \omega^{j_2} , ..., \omega^{j_s}}=J_{\omega^{j_1} , \omega^{j_2} , ..., \omega^{j_s}}/(k_B T).				
			\end{gathered} 
		\end{equation}
		where $ T $ is temperature, $k_B $ is Boltzmann's constant. Let us assume

		 	\begin{equation}\label{Hamiltonian5}
		 	\begin{gathered}
		 \overrightarrow{	\mathcal K}_{\Omega_{\tau^0}}=\{  \{	K_{\omega^{j_1} , \omega^{j_2} , ..., \omega^{j_s}} \},  \{\omega^{j_1},\omega^{j_2},...,\omega^{j_s}\}\subset \Omega_{\tau^0} ,
		 \varnothing < \omega^{j_1} < \omega^{j_2} < ... < \omega^{j_s}
		    \},				
		 	\end{gathered} 
		 \end{equation}
	 Components of the parameter vector of the Hamiltonian $ \overrightarrow{	\mathcal K}_{\Omega_{\tau^0}}$ indexed by all possible non-empty  ordered subsets $\{\omega^{j_1},\omega^{j_2},...,\omega^{j_s}\}\subset \Omega_{\tau^0},$  for example,
	 
	 	\begin{equation}\label{lex_order}
	 	\begin{gathered}
	 	\omega^{0},  \omega^{1},...,\omega^{r}, \{\omega^{0},\omega^{1}\} ,...,  \{  \omega^{0},...,\omega^{r}\}.			
	 	\end{gathered} 
	 \end{equation}
	 
	 	%%%%%%%%%%%%%%%%%%%%%%%%%%%%%%%%%%%%%%%
	 \subsection{Free energy for models with a Hamiltonian on an arbitrary bounded support}\label{Partition_function_arbitrary}

		Partition function can be written in the following form
				
		\begin{equation}\label{Partition_function1} 
			\begin{split}
				Z_{{L}}=\sum_{\sigma}\exp ( -\mathcal {H}(\sigma )/{(k_B T)})=\\
				\sum_{\sigma}\exp ( \sum_{i=0}^{L-1}
					\sum_{\{\omega^{j_1},\omega^{j_2},...,\omega^{j_s}\}\subset \Omega_{\tau^i}}
					K_{\omega^{j_1} , \omega^{j_2} , ...,\omega^{j_s}}\sigma_{\omega^{j_1}}\sigma_{\omega^{j_2}}...\sigma_{\omega^{j_s}}	 ) , 
			\end{split} 
		\end{equation}
	where summation perfomed over all spins.
	
		Let us consider $ {\omega}^{max}=  {\omega}^{r}=\tau^{i_{max}}, $ i.e. $i_{max}$ - the maximum number of a point included in $ \Omega $ ,
		 $ \Omega \subset \{  \tau^0, \tau^1,...,\tau^{i_{max}} \}$.

		To find the partition function we write the elementary transfer matrix $ \Theta=\Theta_{p,q} $ of size $ 2^{i_{max}}  \times 2^{i_{max}} $ in the same way, as in \cite{Khrapov3}, \cite{Khrapov5}.  Nonzero elements of the elementary transfer matrix    $ \Theta=\Theta_{p,q} $ are specified by all sorts of pairs of sets of spins
		$ \{(\sigma_{\tau^0},\sigma_{\tau^{1}},...,\sigma_{\tau^{i_{max}-1}}),$ 
		$(\sigma_{\tau^1},\sigma_{\tau^{2}},...,\sigma_{\tau^{i_{max}}})  \} $:

		\begin{multline} \label{Theta1}
			\Theta_{p,q}=\Theta_{ \{ 
				(\sigma_{\tau^0},\sigma_{\tau^{1}},...,\sigma_{\tau^{i_{max}-1}}),
				(\sigma_{\tau^1},\sigma_{\tau^{2}},...,\sigma_{\tau^{i_{max}}})  \}}=\\
			\exp (
			\sum_{\{\omega^{j_1},\omega^{j_2},...,\omega^{j_s}\}\subset \Omega_{\tau^0}}
			K_{\omega^{j_1} , \omega^{j_2} , ..., \omega^{j_s}}\sigma_{\omega^{j_1}}\sigma_{\omega^{j_2}}...\sigma_{\omega^{j_s}}) ,   
		\end{multline}

		wherein
		\begin{equation}\label{p}
			\begin{split}
			 \Omega_{\tau^0}\subset 
		 \{  \sigma_{\tau^0},\sigma_{\tau^{1}},...,\sigma_{\tau^{i_{max}}} \},\\
				p= \sum_{k=0}^{i_{max}-1}{((1-\sigma_{\tau^k})/2)2^k}, p=0,1,\ldots,2^{i_{max}}-1,	
			\end{split} 
		\end{equation}
		
		\begin{equation}\label{q}
			\begin{split}
				q= \sum_{k=0}^{i_{max}-1}{((1-\sigma_{\tau^{1+k}})/2)2^k}, q=0,1,\ldots,2^{i_{max}}-1.	
			\end{split} 
		\end{equation}

		Then 
		
		\begin{equation}\label{Partition_function3}
			\begin{gathered}
				Z_L=\sum_{ \{\sigma  \}}
				\Theta_{\{ (\sigma_{\tau^0},\sigma_{\tau^{1}},...,\sigma_{\tau^{i_{max}-1}}),
					(\sigma_{\tau^1},\sigma_{\tau^{2}},...,\sigma_{\tau^{i_{max}}})  \} }\\
				\Theta_{\{ (\sigma_{\tau^1},\sigma_{\tau^{2}},...,\sigma_{\tau^{i_{max}}}),
					(\sigma_{\tau^2},\sigma_{\tau^{3}},...,\sigma_{\tau^{i_{max}+1}})  \} }\ldots \\
				\Theta_{\{ (\sigma_{\tau^{L-1}},\sigma_{\tau^{0}},...,\sigma_{\tau^{i_{max}-2}}),
					(\sigma_{\tau^0},\sigma_{\tau^{1}},...,\sigma_{\tau^{i_{max}-1}})  \} } 
				=
				Tr({	\Theta}^L). 
			\end{gathered} 	
		\end{equation}

		By the Perron–Frobenius theorem \cite{Perron} the only one largest eigenvalue $\lambda_{\max}$ of elementary transfer matrix $ \Theta=\Theta_{p,q} $ will correspond to a matrix with positive elements (all matrix elements $ \Theta^{i_{max}} $ will strictly be greater than zero, the structure filled with nonzero elements becomes clear already for $ \Theta^{2} $ . Actually, at first the Perron–Frobenius theorem is used for matrix $ \Theta^{{i_{max}}} $).
		Now the free energy per spin $ f $ in the thermodynamic limit can be written in the following form  \cite{Baxter2016} :
		\begin{equation}\label{free_energy}
			\begin{split}
				f(T, \overrightarrow{	\mathcal K}_{\Omega_{\tau^0}})=-kT \lim\limits_{L\to {\infty}} {\ln(\lambda_{\max}(	\mathcal {L}_{\nu},T,	 \overrightarrow{	\mathcal K}_{\Omega_{\tau^0}}))} , 
			\end{split} 
		\end{equation}
		where $\lambda_{\max}$ is the largest elementary transfer matrix  $ \Theta=\Theta_{p,q} $ eigenvalue.
		  We notice that nonzero elements of matrix 
		  (\ref{Theta1})  $ 	\Theta=	\Theta_{p,q} $  will be the same, if in the set of $(\sigma_{\tau^0},\sigma_{\tau^{1}},...,\sigma_{\tau^{i_{max}}})$ contains the same subset of spins of the support of the Hamiltonian (\ref{Hamiltonian1})  $ \sigma (\Omega_{\tau_0})$. 
		    In general the eigenvector of the elementary transfer matrix (\ref{Theta1})  $	\Theta=	\Theta_{p,q} $, corresponding to the largest eigenvalue  $\lambda_{\max}$ can be represented in the following form:  
		     
		     	\begin{equation}\label{Eigenvector_V}
		     	\begin{gathered}
		     		\overrightarrow{V}=\{  
		     		V_{ (\sigma_{\tau^0},\sigma_{\tau^{1}},...,\sigma_{\tau^{i_{max}-1}})  } , 
		     		\sigma_{\tau^{i} }\in X , i=0,1,...,i_{max}-1 \}=\\
		     		\{ V_p, V_p >0 , 	p= \sum_{k=0}^{i_{max}-1}{((1-\sigma_{\tau^k})/2)2^k}, p=0,1,\ldots,2^{i_{max}-1}-1    \}	.		
		     	\end{gathered} 	
		     \end{equation}
		     
		    We get the system of $2^{i_{max}}$ equations for finding the largest eigenvalue $\lambda_{\max}$ of the elementary transfer matrix 
		    
		    	\begin{equation}\label{main_system}%{Eigenvalue_max}
		    	\begin{gathered}
		    		{\lambda_{max}}	V_{ (\sigma_{\tau^0},\sigma_{\tau^{1}},...,\sigma_{\tau^{i_{max}-1}})  } =\\
		    			\sum_{\sigma_{\tau^{i_{max}}}} 
		    	\Theta_{\{ (\sigma_{\tau^0},\sigma_{\tau^{1}},...,\sigma_{\tau^{i_{max}-1}}),
		    	(\sigma_{\tau^1},\sigma_{\tau^{2}},...,\sigma_{\tau^{i_{max}}})  \} }
		    			V_{ (\sigma_{\tau^1},\sigma_{\tau^{2}},...,\sigma_{\tau^{i_{max}}})},\\\sigma_{\tau^i}	\in X=\{{+1,-1} \}, i=0,1,...,i_{max}.		    			
		    	\end{gathered} 	
		    \end{equation}

		    Then we consider the case when the components 
		    $V_{ (\sigma_{\tau^0},\sigma_{\tau^{1}},...,\sigma_{\tau^{i_{max}-1}})  } $ of eigenvector $ 	\overrightarrow{V}$ (\ref{Eigenvector_V}) depend only on
		     $\sigma_{\phi}, {\phi} \in  \Phi= \Phi (\Omega)  $, and don't depend on
		      $\sigma_{\tau}, {\tau} \notin  \Phi= \Phi (\Omega)  $
		     . I.e.
		     	\begin{equation}\label{Eigenvector_V_Phi}
		     	\begin{gathered}
		     		V_{ (\sigma_{\tau^0},\sigma_{\tau^{1}},...,\sigma_{\tau^{i_{max}-1}})  }=V_{ (\sigma_{\tau^0},\sigma_{\tau^{1}},...,\sigma_{\tau^{i_{max}-1}})  }(\sigma_{\tau^{s_k}}=\sigma_{\phi^{k}}, k=1, 2, ..., m).\\		    			
		     	\end{gathered} 	
		     \end{equation}

		     With this restriction on the form of the eigenvector, the number of equations in the system of equations (\ref{main_system}) can be significantly reduced, and the system of equations itself can be written in the form
		     
		     	\begin{equation}\label{main_system_small}
		     	\begin{gathered}
		     		{\lambda_{max}}	w_{ (\sigma_{\omega^0},\sigma_{\omega^{1}},...,\sigma_{\omega^{r-1}})  }(\sigma_{\omega^{l_k}}=\sigma_{\phi^{k}}, k=1, 2, ..., m) =\\
		     		\sum_{\sigma_{\omega^{r}}} 
		     		\Theta_{\{ (\sigma_{\omega^0},\sigma_{\omega^{1}},...,\sigma_{\omega^{r-1}}),
		     			(\sigma_{\omega^1},\sigma_{\omega^{2}},...,\sigma_{\omega^{r}})  \} }
		     		w_{ (\sigma_{\omega^1},\sigma_{\omega^{2}},...,\sigma_{\omega^{r}})}
		     		(\sigma_{\omega^{l_{k+1}}}=\sigma_{\phi^{k}}, k=1, 2, ..., m)
		     		,		    			
		     	\end{gathered} 	
		     \end{equation}
		     where from (\ref{Phi}) we have
		     	\begin{equation}\label{Eigenvector_w}%{Eigenvalue_max_omega_w}
		     	\begin{gathered}
		     		\overrightarrow{w}=\{
		     		w_{  (\sigma_{\omega^0},\sigma_{\omega^{1}},...,\sigma_{\omega^{r-1}})  }(\sigma_{\omega^{l_k}}=\sigma_{\phi^{k}}, k=1, 2, ..., m) =\\
		     			W_{ (\sigma_{\phi^1},\sigma_{\phi^{2}},...,\sigma_{\phi^{m}})}\}.
		     	\end{gathered} 	
		     \end{equation}

	        Wherein if $\Phi= \varnothing$, than 
	        	\begin{equation}\label{Eigenvector_w_varnothing}
	        	\begin{gathered}
	        		w_{ (\sigma_{\omega^0},\sigma_{\omega^{1}},...,\sigma_{\omega^{r-1}})  } =
	        		W_{0}.\\		     			     			
	        	\end{gathered} 	
	        \end{equation}
        It is convenient to normalize the eigenvector  $ 	\overrightarrow{w}$, taking  $	W_{0}=1.$

		     The system of equations (\ref{main_system_small})  consists only of $ 2^{|\Omega|-1}$
		     equations, all other equations of the system (\ref{main_system}) will coincide with the equations from the system (\ref{main_system_small}) .

	We rewrite (\ref{main_system_small}) in the following form 	
	
		\begin{equation}\label{main_system_small_2}%{Eigenvalue_max_omega2main}
		\begin{gathered}
			{\lambda_{max}}	w_{ (\sigma_{\omega^0},\sigma_{\omega^{1}},...,\sigma_{\omega^{r-1}})  } =\\
			\sum_{\sigma_{\omega^{r}} \in X} 		
		\exp(  \sum_{\{\omega^{j_1},\omega^{j_2},...,\omega^{j_s}\}\subset \Omega_{\tau^0}}
		K_{\omega^{j_1} , \omega^{j_2} , ..., \omega^{j_s}}\sigma_{\omega^{j_1}}\sigma_{\omega^{j_2}}...\sigma_{\omega^{j_s}})
			w_{ (\sigma_{\omega^1},\sigma_{\omega^{2}},...,\sigma_{\omega^{r}})}		
			,\\
			\sigma_{\omega^j}	\in X, j=0,1,...,r-1.	    			
		\end{gathered} 	
	\end{equation}
	
	We substitute $\lambda_{max}$ from from each equation of the system  (\ref{main_system_small_2}
	
		\begin{equation}\label{lambda_max}
		\begin{gathered}
				{\lambda_{max}}=
			\exp(  \sum_{\{\omega^{j_1},\omega^{j_2},...,\omega^{j_s}\}\subset \Omega'_{\tau^0}}
			K_{\omega^{j_1} , \omega^{j_2} , ..., \omega^{j_s}} \sigma_{\omega^{j_1}}\sigma_{\omega^{j_2}}...\sigma_{\omega^{j_s}})\\
				(			\sum_{\sigma_{\omega^{r}} \in X} 		
			\exp(  \sum_{\{\omega^{j_1},\omega^{j_2},...,\omega^{r}\}\subset \Omega_{\tau^0}}
			K_{\omega^{j_1} , \omega^{j_2} , ..., \omega^{r}}
			\sigma_{\omega^{j_1}}\sigma_{\omega^{j_2}}...
			\sigma_{\omega^{r}})\\
			w_{ (\sigma_{\omega^1},\sigma_{\omega^{2}},...,\sigma_{\omega^{r}})}	)
			/ 	w_{ (\sigma_{\omega^0},\sigma_{\omega^{1}},...,\sigma_{\omega^{r-1}})  }
			,\\
			\sigma_{\omega^j}	\in X, j=0,1,...,r-1.	 
	\end{gathered} 	
	\end{equation}
		
	In (\ref{lambda_max}) each equation of the system is numbered by the set $(\sigma_{\omega^0},\sigma_{\omega^{1}},...,\sigma_{\omega^{r-1}})$.

	We substitute $\lambda_{max}$ from the first equation of the system  (\ref{lambda_max})
	with $\sigma_{\omega^0}=+1 ,...,\sigma_{\omega^{{r}-1}}=+1 $

		\begin{equation}\label{LambdaMaxFirst}
		\begin{gathered}
			{\lambda_{max}}=
				\exp(  \sum_{\{\omega^{j_1},\omega^{j_2},...,\omega^{j_s}\}\subset \Omega'_{\tau^0}}
			K_{\omega^{j_1} , \omega^{j_2} , ..., \omega^{j_s}})\\
			(			\sum_{\sigma_{\omega^{r}}} 		
			\exp(  \sum_{\{\omega^{j_1},\omega^{j_2},...,\omega^{r}\}\subset \Omega_{\tau^0}}
			K_{\omega^{j_1} , \omega^{j_2} , ..., \omega^{r}}\sigma_{\omega^{r}})\\
			w_{ (\sigma_{\omega^1}=+1,\sigma_{\omega^{2}}=+1,...,,\sigma_{\omega^{r-1}}=+1,\sigma_{\omega^{r}})})
		/ \\
				w_{ (\sigma_{\omega^0}=+1,\sigma_{\omega^{1}}=+1,...,\sigma_{\omega^{r-1}}=+1)  }
			.		    			
		\end{gathered} 	
	\end{equation}
	
	and substitute in all the remaining equations of the system (\ref{lambda_max}). We get

		\begin{equation}\label{K_system}
		\begin{gathered}
			\exp(  \sum_{\{\omega^{j_1},\omega^{j_2},...,\omega^{j_s}\}\subset \Omega'_{\tau^0}}
		K_{\omega^{j_1} , \omega^{j_2} , ..., \omega^{j_s}} (1-\sigma_{\omega^{j_1}}\sigma_{\omega^{j_2}}...\sigma_{\omega^{j_s}}))=\\
		\{		(			\sum_{\sigma_{\omega^{r}}} 		
			\exp(  \sum_{\{\omega^{j_1},\omega^{j_2},...,\omega^{r}\}\subset \Omega_{\tau^0}}
			K_{\omega^{j_1} , \omega^{j_2} , ..., \omega^{r}}
			\sigma_{\omega^{j_1}}\sigma_{\omega^{j_2}}...
			\sigma_{\omega^{r}})
			w_{ (\sigma_{\omega^1},\sigma_{\omega^{2}},...,\sigma_{\omega^{r}})}
		)
			/ \\
			(			\sum_{\sigma_{\omega^{r}}} 		
			\exp(  \sum_{\{\omega^{j_1},\omega^{j_2},...,\omega^{r}\}\subset \Omega_{\tau^0}}
			K_{\omega^{j_1} , \omega^{j_2} , ..., \omega^{r}}\sigma_{\omega^{r}})
			w_{ (\sigma_{\omega^1}=+1,\sigma_{\omega^{2}}=+1,...,,\sigma_{\omega^{r-1}}=+1,\sigma_{\omega^{r}})}) \} \\
			(	w_{ (\sigma_{\omega^0}=+1,\sigma_{\omega^{1}}=+1,...,\sigma_{\omega^{r-1}}=+1)  }/ 	
			w_{ (\sigma_{\omega^0},\sigma_{\omega^{1}},...,\sigma_{\omega^{r-1}})  })
			.
		\end{gathered} 	
	\end{equation}
	 
	Let us logarithm the system of equations (\ref{K_system}) and write in the following form 
	
		\begin{multline} \label{K_system_2}%{main_system2_6}
		\sum_{\{\omega^{j_1},\omega^{j_2},...,\omega^{j_s}\}\subset \Omega'_{\tau^0}}
		K_{\omega^{j_1} , \omega^{j_2} , ..., \omega^{j_s}} (1-\sigma_{\omega^{j_1}}\sigma_{\omega^{j_2}}...\sigma_{\omega^{j_s}})
		=
		A_{\sigma_{\omega^0} ,\sigma_{\omega^1},...,\sigma_{\omega^{r-1}} } 
	\end{multline}
	
	for all different values ${\sigma_{\omega^0} ,\sigma_{\omega^1},...,\sigma_{\omega^{r-1}} }$, excluding the set
	${\sigma_{\omega^j} =+1, j=0,1,...,{r-1}.}$
	
	  In the system of equations (\ref{K_system_2})  
		\begin{equation}\label{A_right}
		\begin{gathered}
		 A_{\sigma_{\omega^0} ,\sigma_{\omega^1},...,\sigma_{\omega^{r-1}} }=\\
			\ln(			\sum_{\sigma_{\omega^{r}}} 		
			\exp(  \sum_{\{\omega^{j_1},\omega^{j_2},...,\omega^{r}\}\subset \Omega_{\tau^0}}
			K_{\omega^{j_1} , \omega^{j_2} , ..., \omega^{r}}
			\sigma_{\omega^{j_1}}\sigma_{\omega^{j_2}}...
			\sigma_{\omega^{r}})
			w_{ (\sigma_{\omega^1},\sigma_{\omega^{2}},...,\sigma_{\omega^{r}})}
			)
			- \\
			\ln(			\sum_{\sigma_{\omega^{r}}} 		
			\exp(  \sum_{\{\omega^{j_1},\omega^{j_2},...,\omega^{r}\}\subset \Omega_{\tau^0}}
			K_{\omega^{j_1} , \omega^{j_2} , ..., \omega^{r}}\sigma_{\omega^{r}})
			w_{ (\sigma_{\omega^1}=+1,\sigma_{\omega^{2}}=+1,...,,\sigma_{\omega^{r-1}}=+1,\sigma_{\omega^{r}})})  +\\
			\ln(	w_{ (\sigma_{\omega^0}=+1,\sigma_{\omega^{1}}=+1,...,\sigma_{\omega^{r-1}}=+1)  })- 	
			\ln(w_{ (\sigma_{\omega^0},\sigma_{\omega^{1}},...,\sigma_{\omega^{r-1}})  } )
			.
		\end{gathered} 	
	\end{equation}

   We notice that  

		\begin{equation}\label{A1}
		\begin{gathered}
			A_{\sigma_{\omega^0}=+1 ,\sigma_{\omega^1}=+1,...,\sigma_{\omega^{r-1}}=+1 }=0.
		\end{gathered} 	
	\end{equation} 
	
	The system of linear  equations (\ref{K_system_2}) was received by the set of coefficients 
	$\{  	K_{\omega^{j_1} , \omega^{j_2} , ..., \omega^{j_s}},  
	{\{\omega^{j_1} , \omega^{j_2} , ..., \omega^{j_s}\}\subset \Omega'_{\tau^0}}    \}$.
	
	Let's rewrite it in matrix form
	
		\begin{multline} \label{GKA}
		G	 \overrightarrow{	\mathcal K}_{\Omega'_{\tau^0}}= \overrightarrow{ A},\\
	\end{multline}
   where
   	\begin{equation}\label{G}
   	\begin{gathered}
   	G=\{G(\{ {\sigma_{\omega^0} ,\sigma_{\omega^1},...,\sigma_{\omega^{r-1}} }\};  
   	\{{\omega^{j_1} , \omega^{j_2} , ..., \omega^{j_s}}\} \subset \Omega'_{\tau^0}   )\} ,\\
   	G(\{ {\sigma_{\omega^0} ,\sigma_{\omega^1},...,\sigma_{\omega^{r-1}} }\};  
   	\{{\omega^{j_1} , \omega^{j_2} , ..., \omega^{j_s}}\} \subset \Omega'_{\tau^0}   )=\\ (1-\sigma_{\omega^{j_1}}\sigma_{\omega^{j_2}}...\sigma_{\omega^{j_s}}), 			
   	\end{gathered} 	
   \end{equation}

	\begin{equation}\label{A_vector}
	\begin{gathered}
		\overrightarrow{ A}=\{A_{\sigma_{\omega^0} ,\sigma_{\omega^1},...,\sigma_{\omega^{r-1}} } =A_p,
		p= \sum_{k=0}^{r-1}{((1-\sigma_{\omega^k})/2)2^k}, p=0,1,\ldots,2^{r}-1\}.
	\end{gathered} 	
\end{equation}
	
	The matrix elements of the matrix $G^{-1}$ have the form 
	
		\begin{equation}\label{G_1}
		\begin{gathered}
		 G^{-1}(  
		\{{\omega^{j_1} , \omega^{j_2} , ..., \omega^{j_s}}\} \subset \Omega'_{\tau^0};\{ {\sigma_{\omega^0} ,\sigma_{\omega^1},...,\sigma_{\omega^{r-1}} }\}   ) =\\ -\sigma_{\omega^{j_1}}\sigma_{\omega^{j_2}}...\sigma_{\omega^{j_s}} /{2^{r}}. 			
		\end{gathered} 	
	\end{equation}

	This implies that
	
		\begin{equation}\label{K}
		\begin{gathered}			
			K_{\omega^{j_1} , \omega^{j_2} , ..., \omega^{j_s}}=\\
			 -(\sum_{\{\sigma_{\omega^0},\sigma_{\omega^1},...,\sigma_{\omega^{r-1}}\}}
			\sigma_{\omega^{j_1}}\sigma_{\omega^{j_2}}...\sigma_{\omega^{j_s}}	
					A_{\sigma_{\omega^0} ,\sigma_{\omega^1},...,\sigma_{\omega^{r-1}} })/2^{r},\\				
					\{\sigma_{\omega^0} ,\sigma_{\omega^1},...,\sigma_{\omega^{r-1}} \} 	\neq
					(+1,+1,...,+1 )				 . 			
		\end{gathered} 	
	\end{equation}

    Free energy $f$ can be found from the following ratio
    	\begin{equation}\label{free_energy2}
    	\begin{gathered}
    		- f/ {(kT)}= \ln (	\lambda_{max}),
    	\end{gathered} 	
    \end{equation}
    where
    $	\lambda_{max} $ can be found from (\ref{LambdaMaxFirst}) .

	%%%%%%%%%%%%%%%%%%%%%%%%%%%%%%%%%%%%%%%
		\subsection{Free energy for models with the Hamiltonians invariant under the reversal of signs of all spins}\label{even} 
	
	  In our case, the Hamiltonian of such models has the form (\ref{Hamiltonian1}), where  $s=2 p $ are only even.
	  This halves the number of non-zero coefficients $	K_{\omega^{j_1} , \omega^{j_2} , ..., \omega^{j_s}}.$ But this also halves the number of equations of the system   (\ref{main_system_small_2}): with such Hamiltonians the elementary transfer matrix (\ref{Theta1}) and the  eigenvectors (\ref{Eigenvector_w}) are centrally symmetric:

	  	\begin{equation}\label{Teta_even}
	  	\begin{gathered}			
	  		\Theta_{p,q}=\Theta_{ \{ 
	  		(\sigma_{\tau^0},\sigma_{\tau^{1}},...,\sigma_{\tau^{i_{max}-1}}),
	  		(\sigma_{\tau^1},\sigma_{\tau^{2}},...,\sigma_{\tau^{i_{max}}})  \}}=\\
\Theta_{ \{ 
	(-\sigma_{\tau^0},-\sigma_{\tau^{1}},...,-\sigma_{\tau^{i_{max}-1}}),
	(-\sigma_{\tau^1},-\sigma_{\tau^{2}},...,-\sigma_{\tau^{i_{max}}})  \}} , 			
	  	\end{gathered} 	
	  \end{equation}
	  
	  	\begin{equation}\label{w_even}
	  	\begin{gathered}
	  		w_{ (\sigma_{\omega^0},\sigma_{\omega^{1}},...,\sigma_{\omega^{r-1}})  } =
	  		w_{ (-\sigma_{\omega^0},-\sigma_{\omega^{1}},...,-\sigma_{\omega^{r-1}})  }.\\		     			     			
	  	\end{gathered} 	
	  \end{equation}
	    
	    Therefore, it is enough to leave the equations in the system of equations (\ref{main_system_small_2}) in which
	     $\sigma_{\omega^{r-1}}=+1$.	  

	 We take arbitrary interaction coefficients as free parameters  
	  $	K_{\omega^{j_1} , \omega^{j_2} , ..., \omega^{j_{2 p - 1}}, \omega^{r}}$ including the point $\omega^{r}$ and corresponding to the interaction of an even number of spins, and we take the interaction coefficients 
	  $	K_{\omega^{j_1} , \omega^{j_2} , ..., \omega^{j_{2 p}}, \omega^{r}}$ including the point $\omega^{r}$ and corresponding to the interaction of an odd number of spins, after this we equate them to zero: 
	  $	K_{\omega^{j_1} , \omega^{j_2} , ..., \omega^{j_{2 p}}, \omega^{r}}=0$. In (\ref{lambda_max})  each equation of the system is numbered by the set $(\sigma_{\omega^0},\sigma_{\omega^{1}},...,\sigma_{\omega^{r-1}})$. If we change all the spin signs in the set 
$(\sigma_{\omega^0},\sigma_{\omega^{1}},...,\sigma_{\omega^{r-1}})$, then $	{\lambda_{max}}$ and the formula

	\begin{equation}\label{sum} 
		\begin{gathered}		
			(			\sum_{\sigma_{\omega^{r} \in X}} 		
			\exp(  \sum_{\{\omega^{j_1},\omega^{j_2},...,\omega^{r}\}\subset \Omega_{\tau^0}}
			K_{\omega^{j_1} , \omega^{j_2} , ..., \omega^{r}}
			\sigma_{\omega^{j_1}}\sigma_{\omega^{j_2}}...
			\sigma_{\omega^{r}})\\
			w_{ (\sigma_{\omega^1},\sigma_{\omega^{2}},...,\sigma_{\omega^{r}})}	)
			/ 	w_{ (\sigma_{\omega^0},\sigma_{\omega^{1}},...,\sigma_{\omega^{r-1}})  }
			,\\
			\sigma_{\omega^j}	\in X, j=0,1,...,r-1,	 
		\end{gathered} 	
	\end{equation} 

	from  (\ref{lambda_max}) will not change. This is true for any set $(\sigma_{\omega^0},\sigma_{\omega^{1}},...,\sigma_{\omega^{r-1}})$. Consequently, and the remaining formula 
		from  (\ref{lambda_max}) 
		
	\begin{equation}\label{sum2} 
		\begin{gathered}			
		 \sum_{\{\omega^{j_1},\omega^{j_2},...,\omega^{j_s}\}\subset \Omega'_{\tau^0}}
			K_{\omega^{j_1} , \omega^{j_2} , ..., \omega^{j_s}} \sigma_{\omega^{j_1}}\sigma_{\omega^{j_2}}...\sigma_{\omega^{j_s}}
			,\\
			\sigma_{\omega^j}	\in X, j=0,1,...,r-1,	 
		\end{gathered} 	
	\end{equation}
	will not change when changing the signs of all spins in the sets
	$(\sigma_{\omega^0},\sigma_{\omega^{1}},...,\sigma_{\omega^{r-1}})$ for any such set. This implies that coefficients  
	$	K_{\omega^{j_1} , \omega^{j_2} , ..., \omega^{j_{2 p -1}}}=0$ for an odd number of spins in the interaction. And we can use the general formula (\ref{K}) taking as free parameters only the interaction coefficients
	$	K_{\omega^{j_1} , \omega^{j_2} , ..., \omega^{j_{2 p - 1}}, \omega^{r}}$ including the point $\omega^{r}$ and corresponding to the interaction of an even number of spins and centrally symmetric vector coordinates (\ref{w_even}). In this case, we automatically obtain	$	K_{\omega^{j_1} , \omega^{j_2} , ..., \omega^{j_{2 p -1}}}=0$ for ${\{\omega^{j_1},\omega^{j_2},...,\omega^{j_{2 p -1}}\}\subset \Omega'_{\tau^0}} $ .
	 Taking into account that for the right-hand side (\ref{A_right}) of the system of equations (\ref{K_system_2})  the following equality is true
	 	   
\begin{equation}\label{AA_even}
	\begin{gathered}			
	A_{\sigma_{\omega^0} ,\sigma_{\omega^1},...,\sigma_{\omega^{r-1}} }=	A_{-\sigma_{\omega^0} ,-\sigma_{\omega^1},...,-\sigma_{\omega^{r-1}} } 	  				
	\end{gathered} 	
\end{equation}

	   we can simplify the formula (\ref{K}) for the case, when the Hamiltonians have only nonzero interaction coefficients of the product of an even number of spins and centrally symmetric eigenvector (\ref{w_even}):
	
	  	\begin{equation}\label{K_even}
	  	\begin{gathered}			
	  		K_{\omega^{j_1} , \omega^{j_2} , ..., \omega^{j_{2 p}}}=\\
	  		-(\sum_{\{\sigma_{\omega^0},\sigma_{\omega^1},...,\sigma_{\omega^{r-2}}\}}
	  		\sigma_{\omega^{j_1}}\sigma_{\omega^{j_2}}...\sigma_{\omega^{j_{2 p}}}	
	  		A_{\sigma_{\omega^0} ,\sigma_{\omega^1},...,\sigma_{\omega^{r-2}},
	  			\sigma_{\omega^{r-1}}=+1 })/2^{r-1},\\  			
	  		\{\sigma_{\omega^0} ,\sigma_{\omega^1},...,\sigma_{\omega^{r-1}} \} 	\neq
	  		(+1,+1,...,+1 )	,
	  			{\{\omega^{j_1},\omega^{j_2},...,\omega^{j_{2 p}}\}\subset \Omega'_{\tau^0}},\\ 	
	  			K_{\omega^{j_1} , \omega^{j_2} , ..., \omega^{j_{2 p-1}}}=	0,\\
	  			{\{\omega^{j_1},\omega^{j_2},...,\omega^{j_{2 p -1}}\}\subset \Omega'_{\tau^0}},	  				
	  	\end{gathered} 	
	  \end{equation}
	  
	where   $	A_{\sigma_{\omega^0} ,\sigma_{\omega^1},...,\sigma_{\omega^{r-2}},
		\sigma_{\omega^{r-1}}=+1 } $ we find by the formulas (\ref{A_right}) with a centrally symmetric vector $	\overrightarrow{ w}$ (\ref{w_even}).
		
	   Let us consider various special cases now.
	
	%%%%%%%%%%%%%%%%%%%%%%%%
		\section{Free energy  for $\nu$ - dimensional models with support of the Hamiltonian in the form of  $\nu$ - dimensional simplex }\label{nuD_simplex}
			
		Let us consider unit $\nu$ - dimensional simplex as $\Omega$:
			
		\begin{equation}\label{simplex_nu}
			\begin{gathered}  
				\mathcal {S}_{\nu}=\{t=(t_1,t_2,...,t_{\nu}) \in\mathcal {L}_{\nu} : t_i=0,1, i=1,2,...,\nu\}=\\		
				=\{{\omega}^0, \omega^1,...,\omega^{\nu} \} , \\
				{\omega}^0=\tau^0,  	{\omega}^1=\tau^1, 	{\omega}^2=\tau^{L_1},
				{\omega}^3=\tau^{L_1 L_2}, 	{\omega}^4=\tau^{L_1 L_2 L_3}, ...,\\
				{\omega}^{\nu}=\tau^{L_1 L_2...L_{\nu-1}}= 		{\omega}^{max}={\omega}^{r}.
			\end{gathered} 
		\end{equation}
		
		Then

		\begin{equation}\label{phi_simplex_nu}
			\begin{gathered}
				\Phi =\{{\omega}^0\} .
			\end{gathered} 
		\end{equation}
				
		The Hamiltonian of the model has the form  (\ref{Hamiltonian1}) ,		 
		where $\tau^i=(\tau_1^i,\tau_2^i,...,\tau_{\nu}^i) \in  \mathcal {L}_{\nu} , \;\; \Omega=	\mathcal {S}_{\nu} , r=\nu.  $

		Consequently,

		\begin{equation}\label{w_simplex_nu}
			\begin{gathered}
				w_{ (\sigma_{\omega^0},\sigma_{\omega^{1}},...,\sigma_{\omega^{{\nu-1}}})  }(\sigma_{\omega^{0}}=\sigma_{\phi^{1}}) =
				W_{ (\sigma_{\phi^1})}.\\		     			     			
			\end{gathered} 	
		\end{equation}
				
		It means that the eigenvector of the elementary transfer matrix $ \Theta=\Theta_{p,q} $, corresponding to
		the largest eigenvalue $\lambda_{max} $, we represent in the following form : \\
		
		\begin{equation}\label{nuD_eigenvector}   
			\overrightarrow{V}=
			(\underbrace{W_0,W_1,\dots,W_0,W_1 }_{2^{L_1 L_2 ... L_{\nu-1}}})^T
			.
		\end{equation}	
		 
		As a normalization of the eigenvector  $	\overrightarrow{V}$ is convenient to take $W_0=1.$
		
		In the solution (\ref{K}) for the case of $\nu$-dimensional simplex we need to take $\Omega=	\mathcal {S}_{\nu}$ ,  $\Omega'=	\mathcal {S}'_{\nu}=\{{\omega}^0, \omega^1,...,\omega^{\nu-1} \} $, $r=\nu .$ Then

		\begin{equation}\label{K_nuD_simplex}
			\begin{gathered}			
				K_{\omega^{j_1} , \omega^{j_2} , ..., \omega^{j_s}}=\\
				-(\sum_{\{\sigma_{\omega^0},\sigma_{\omega^1},...,\sigma_{\omega^{\nu-1}}\}}
				\sigma_{\omega^{j_1}}\sigma_{\omega^{j_2}}...\sigma_{\omega^{j_s}}	
				A_{\sigma_{\omega^0} ,\sigma_{\omega^1},...,\sigma_{{\omega^{\nu-1}}} })/2^{{\nu}},\\
				\{\sigma_{\omega^0} ,\sigma_{\omega^1},...,\sigma_{\omega^{{\nu-1}}} \} 	\neq
				(+1,+1,...,+1 )	, 			
			\end{gathered} 	
		\end{equation}

		where 
		
		\begin{equation}\label{A_nuD_simplex}
			\begin{gathered}
				A_{\sigma_{\omega^0} ,\sigma_{\omega^1},...,\sigma_{\omega^{\nu-1} }}=\\
				\ln(			\sum_{\sigma_{\omega^{{\nu}}}} 		
				\exp(  \sum_{\{\omega^{j_1},\omega^{j_2},...,\omega^{{\nu}}\}\subset \Omega_{\tau^0}}
				K_{\omega^{j_1} , \omega^{j_2} , ..., \omega^{{\nu}}}
				\sigma_{\omega^{j_1}}\sigma_{\omega^{j_2}}...
				\sigma_{\omega^{\nu}}) \cdot \\		
				w_{ (\sigma_{\omega^{1}},\sigma_{\omega^{2}},...
					,
					\sigma_{\omega^{{\nu-1}}},\sigma_{\omega^{{\nu}}})  }
				)
				- \\
				\ln(			\sum_{\sigma_{\omega^{{\nu}}}} 		
				\exp(  \sum_{\{\omega^{j_1},\omega^{j_2},...,\omega^{{\nu}}\}\subset \Omega_{\tau^0}}
				K_{\omega^{j_1} , \omega^{j_2} , ..., \omega^{{\nu}}}\sigma_{\omega^{{\nu}}}) \cdot \\
				w_{ (\sigma_{\omega^1}=+1,\sigma_{\omega^{2}}=+1,...,\sigma_{\omega^{{\nu-1}}}=+1,\sigma_{\omega^{{\nu}}})})  +\\
				\ln(	w_{ (\sigma_{\omega^0}=+1,\sigma_{\omega^{1}}=+1,...,\sigma_{\omega^{{\nu-1}}}=+1)  })- 	
				\ln(w_{ (\sigma_{\omega^0},\sigma_{\omega^{1}},\sigma_{\omega^{2}},...,	\sigma_{\omega^{{\nu-1}}})  }  ).
			\end{gathered} 	
		\end{equation}
				
		By revising $\nu$ - dimensional models with the support of the Hamiltonian in the form of  $\nu$ - dimensional  simplex containing nonzero only the coefficients corresponding to the interaction of an even number of spins we use the formulas (\ref{K_even}). The eigenvector $	\overrightarrow{V}$ should look like 	
		(\ref{w_simplex_nu}),	(\ref{nuD_eigenvector}) and satisfy the central symmetry condition (\ref{w_even}): 
		\begin{equation}\label{nuD_simplex_even_eigenvector}   
			\overrightarrow{V}=
			(\underbrace{W_0,W_0,\dots,W_0,W_0 }_{2^{L_1 L_2 ... L_{\nu-1}}})^T
			.
		\end{equation}	
	
	  Let us normalize $\overrightarrow{V}$ , putting $W_0=1$. Then for $\nu$ - dimensional models with the support of the Hamiltonian in the form of $\nu$ - dimensional simplex containing nonzero only the coefficients corresponding to the interaction of an even number of spins the formulas (\ref{K_nuD_simplex}), (\ref{A_nuD_simplex})  simplify
	  	  
	  	\begin{equation}\label{K_nuD_simplex_even}
	  	\begin{gathered}			
	  		K_{\omega^{j_1} , \omega^{j_2} , ..., \omega^{j_{2 p}}}=\\
	  		-(\sum_{\{\sigma_{\omega^0},\sigma_{\omega^1},...,\sigma_{\omega^{{\nu}-2}}\}}
	  		\sigma_{\omega^{j_1}}\sigma_{\omega^{j_2}}...\sigma_{\omega^{j_{2 p}}}	
	  		A_{\sigma_{\omega^0} ,\sigma_{\omega^1},...,\sigma_{\omega^{{\nu}-2}},
	  			\sigma_{\omega^{{\nu}-1}}=+1 })/2^{{\nu}-1},\\
	  		\{\sigma_{\omega^0} ,\sigma_{\omega^1},...,\sigma_{\omega^{{\nu}-1}} \} 	\neq
	  		(+1,+1,...,+1 )	,
	  		{\{\omega^{j_1},\omega^{j_2},...,\omega^{j_{2 p}}\}\subset \Omega'_{\tau^0}},\\ 	
	  		K_{\omega^{j_1} , \omega^{j_2} , ..., \omega^{j_{2 p-1}}}=	0,\\
	  		{\{\omega^{j_1},\omega^{j_2},...,\omega^{j_{2 p -1}}\}\subset \Omega'_{\tau^0}},	  				
	  	\end{gathered} 	
	  \end{equation}
	   where
	  
	  	\begin{equation}\label{A_nuD_simplex_even}
	  	\begin{gathered}
	  		A_{\sigma_{\omega^0} ,\sigma_{\omega^1},...,\sigma_{\omega^{\nu-1} }}=\\
	  		\ln(			\sum_{\sigma_{\omega^{{\nu}}}} 		
	  		\exp(  \sum_{\{\omega^{j_1},\omega^{j_2},...,\omega^{{\nu}}\}\subset \Omega_{\tau^0}}
	  		K_{\omega^{j_1} , \omega^{j_2} , ..., \omega^{{\nu}}}
	  		\sigma_{\omega^{j_1}}\sigma_{\omega^{j_2}}...
	  		\sigma_{\omega^{\nu}}))
	  		- \\
	  		\ln(			\sum_{\sigma_{\omega^{{\nu}}}} 		
	  		\exp(  \sum_{\{\omega^{j_1},\omega^{j_2},...,\omega^{{\nu}}\}\subset \Omega_{\tau^0}}
	  		K_{\omega^{j_1} , \omega^{j_2} , ..., \omega^{{\nu}}}\sigma_{\omega^{{\nu}}})) .
	  	\end{gathered} 	
	  \end{equation}

	   Next, consider in detail as examples lattice models with the support of the Hamiltonian in the form of simplex for $\nu=2$ and $\nu=3.$
		
		%%%%%%%%%%%%%%%%%%%%%%%%%%%%%%%%%%%%%%%%%%%%%%%%%%%%%%%%%%%%%%%%%%%%%%%%%%%%%%%%%%%%%%%%%%%
		\subsection{Free energy of the "checkerboard-triangular" model on the triangular lattice }\label{checkerboard-triangular}
	
		Let us assume $\nu=2.$  The Hamiltonian of the "checkerboard-triangular" model on the triangular lattice (Fig.1) has the form 
		\begin{equation}\label{Chess_Hamiltonian1}
			\begin{gathered}
				\mathcal {H} (\sigma )= -\sum_{i=0,{\omega^{j}\subset \Omega_{\tau^i}}, j=0,1,2}^{L - 1} 		    	
				(	K_{\omega^{0}} 	\sigma_{\omega^{0}}+
				K_{\omega^{1}} 	\sigma_{\omega^{1}}+	K_{\omega^{2}} 	\sigma_{\omega^{2}}+\\
				K_{\omega^{0} \omega^{1} } \sigma_{\omega^{0}} \sigma_{\omega^{1}}+
				K_{\omega^{0} \omega^{2} } \sigma_{\omega^{0}} \sigma_{\omega^{2}}+
				K_{\omega^{1} \omega^{2} } \sigma_{\omega^{1}} \sigma_{\omega^{2}}+
				K_{\omega^{0} \omega^{1} \omega^{2} } \sigma_{\omega^{0}}  \sigma_{\omega^{1}} \sigma_{\omega^{2}}	).
			\end{gathered} 
		\end{equation}

		\begin{picture}(180,180) 		
			\put(20,55){\line(1,0){100}}
			\put(20,55){\line(0,1){100}}
			\put(20,55){\circle*{3}}
			\put(120,55){\circle*{3}}
			\put(0,40){${\tau^0}={\omega^0}$}
			\put(100,40){${\tau^1}={{\omega^1}}$}	
			\put(215,40){${\tau^2}$}
			\put(20,155){\line(1,0){100}}
			\put(220,55){\circle*{3}}		
			\put(120,55){\line(0,1){100}}
			\put(120,55){\line(-1,1){100}}
			\put(120,55){\line(1,0){100}}
			\put(100,55){\line(-1,1){80}}
			\put(80,55){\line(-1,1){60}}
			\put(60,55){\line(-1,1){40}}
			\put(40,55){\line(-1,1){20}}		
			\put(20,155){\circle*{3}}
			\put(120,155){\circle*{3}}
			\put(0,165){${\tau^{L_1}}={\omega^2}$}
			\put(100,165){${\tau^{L_1 +1}}={\omega^3}$}		
			\qbezier[460](220,55)(520,90)(20,155)
			\put(55,10){Fig. 1}
		\end{picture}
				
		From  \ref{simplex_nu})  and  \ref{phi_simplex_nu}) we have $\Omega= \{ \omega^0,  \omega^1,  \omega^2 \},$ $r=2$,
		$\Phi=\{\phi^1= \omega^0 \}$.  Consequently,
		\begin{equation}\label{Chess_eigenvector_1}
			\begin{gathered}
				w_{ (\sigma_{\omega^0},\sigma_{\omega^{1}})  }=w_{ (\sigma_{\omega^0},\sigma_{\omega^{1}})  }(\sigma_{\omega^{0}}=\sigma_{\phi^{1}}) =
				W_0 (\sigma_{\omega^{0}} + 1)/2 + W_1 (1 - \sigma_{\omega^{0}})/2, \\
				w_{ (\sigma_{\omega^1},\sigma_{\omega^{2}})  }(\sigma_{\omega^{1}}=\sigma_{\phi^{1}}) =\\
				W_0 (\sigma_{\omega^{1}} + 1)/2 +  W_1 (1 - \sigma_{\omega^{1}})/2. \\					    			
			\end{gathered} 	
		\end{equation}
		Here $ W_{\sigma_{\phi^1}}=	W_0 (\sigma_{\phi^{1}} + 1)/2 + W_1 (1 - \sigma_{\phi^{1}})/2 .$   
		
		It is convenient for us to normalize the eigenvector by setting $W_0=1.$       
		
		It means that the eigenvector of the elementary transfer matrix $ \Theta=\Theta_{p,q} $, corresponding to
		the largest eigenvalue $ \lambda_{max}$, we represent in the following form: \\
		
		\begin{equation}\label{Chess_eigenvector_2}   
			\overrightarrow{V}=
			(\underbrace{W_0,W_1,\dots,W_0,W_1 }_{2^{L_1}})^T
			.
		\end{equation}	
		
		 Further, in subsubsection \ref{checkerboard-triangular-main}, we consider the general case with the Hamiltonian (\ref{Chess_Hamiltonian1}), in subsubsection \ref{checkerboard-triangular-paired} with the Hamiltonian containing only pair interactions.
		 
		 \subsubsection{ "Checkerboard-triangular" model, general case}\label{checkerboard-triangular-main}
		
		Let us write the system of 4 equations (\ref{main_system_small}) for "checkerboard-triangular case" (other $2^{L_1} - 4 $ equations of the system  (\ref{main_system}) will coincide one of these 4 equations).
			
		\begin{equation}\label{Chess_Eigenvalue_max_omega}
			\begin{gathered}
				{\lambda_{max}}	w_{ (\sigma_{\omega^0},\sigma_{\omega^{1}})  } =
				\sum_{\sigma_{\omega^{2}}} 
				\Theta_{\{ (\sigma_{\omega^0},\sigma_{\omega^{1}}),
					(\sigma_{\omega^1},\sigma_{\omega^{2}})  \} }
				w_{ (\sigma_{\omega^1},\sigma_{\omega^{2}})}
				,		\\
				\sigma_{\omega^0} \in X=\{+1,-1 \},\sigma_{\omega^{1}}	 \in X=\{+1,-1 \}	.
			\end{gathered} 	
		\end{equation}
					
		Then from (\ref{K_nuD_simplex}) we have
		\begin{equation}\label{Chess_K}
			\begin{gathered}
				K_{\omega^{0}}=-1/4 \;\; ((-1) 	A_{ (-1,+1) } +(+1) 	A_{ (+1,-1) } + 
				(-1) 	A_{ (-1,-1) })  \\
				K_{\omega^{1}}=-1/4 \;\; ((+1) 	A_{ (-1,+1) }+(-1) 	A_{ (+1,-1) } +
				(-1) 	A_{ (-1,-1) })  \\		
				K_{\omega^{0} \omega^{1}}=-1/4 \;\; ((-1) (+1) 	A_{ (-1,+1) }+
				(+1) (-1) 	A_{ (+1,-1) } +\\
				(-1) (-1)	A_{ (-1,-1) }),
			\end{gathered} 	
		\end{equation}
		where from (\ref{A_nuD_simplex}), (\ref{A_right})
		 
		\begin{equation}\label{Chess_A}
			\begin{gathered}
				A_{ (\sigma_{\omega^0},\sigma_{\omega^{1}}) }
				=
				\ln(			\sum_{\sigma_{\omega^{2}}} 		
				\exp(	K_{\omega^{2}} 	\sigma_{\omega^{2}}+
				K_{\omega^{0} \omega^{2} } \sigma_{\omega^{0}} \sigma_{\omega^{2}}+
				K_{\omega^{1} \omega^{2} } \sigma_{\omega^{1}} \sigma_{\omega^{2}}+\\
				K_{\omega^{0} \omega^{1} \omega^{2} } \sigma_{\omega^{0}}  \sigma_{\omega^{1}} \sigma_{\omega^{2}}	)
				w_{ (\sigma_{\omega^1},\sigma_{\omega^{2}})}
				)
				- \\
				\ln(			\sum_{\sigma_{\omega^{2}}} 		
				\exp(	K_{\omega^{2}} 	\sigma_{\omega^{2}}+
				K_{\omega^{0} \omega^{2} }  \sigma_{\omega^{2}}+
				K_{\omega^{1} \omega^{2} }  \sigma_{\omega^{2}}+
				K_{\omega^{0} \omega^{1} \omega^{2} } \sigma_{\omega^{2}}	)
				w_{ (\sigma_{\omega^1}=+1,\sigma_{\omega^{2}})})		 		
				+\\
				\ln(	w_{ (\sigma_{\omega^0}=+1,\sigma_{\omega^{1}}=+1)  })- 	
				\ln(w_{ (\sigma_{\omega^0},\sigma_{\omega^{1}})  } )
				,\\
				(\sigma_{\omega^0},\sigma_{\omega^{1}}) \in \{ (-1,+1),(+1,-1),(-1,-1)\}	.	
			\end{gathered} 	
		\end{equation}
				
		\textbf{Example.} {"Checkerboard-triangular" lattice}.
						
		We define the free parameters:
		
		$W_0=1, W_1=2, K_{\omega^2}=1.2, K_{\omega^0, \omega^{2}}=-2.02, 
		{ K_{\omega^1, \omega^{2}}=-1.12 },{ K_{\omega^0, \omega^1, \omega^{2}}=3.012.} $ 
		
		Other parameters are calculated with (\ref{Chess_K}) ,  (\ref{Chess_A})  :
		\begin{equation}\label{Chess_calculate}
			\begin{gathered}
				K_{\omega^0}= 0.7820255093754221, K_{\omega^1}=2.3025685587397597, \\ 
				{ K_{\omega^0, \omega^{1}}=-1.1892012158890553 }, 
				\lambda_{max}= 21.719383860264113 , \\
				- f/ {(kT)}= 3.0782051272722186 .
			\end{gathered} 	
		\end{equation}
		where $f$ is free energy per one lattice site.
		
		At  the "checkerboard-triangular" lattice let us compare the obtained results with the results from \cite{Wu1985}. Below we will add the top or bottom index $W$ to the notations from \cite{Wu1985} in order not to confuse then with the notation of this paper. Let us write the formalas for calculating $F_W$ from \cite{Wu1985} in a convenient for comparison form.
				
		\begin{equation}\label{Wu2_case1_1}
			\begin{gathered}
				A_W = \sinh(2 (K_3^W + K^W)), \;\;   C_W = \sinh(2 (K_3^W - K^W)),   \\
				B_W = \exp(2 K_3^W) \cosh(2 (K_1^W + K_2^W)) - \exp(-2 K_3^W) \cosh(2 (K_1^W - K_2^W)),\\  
			\end{gathered} 	
		\end{equation}

		\begin{multline} \label{Wu3_case1_1}        \\
			t_1^W = (-B_W + \sqrt{B_W*B_W - 4 A_W C_W})/(2 A_W),\\
			t_2^W = (-B_W - \sqrt{B_W*B_W - 4 A_W C_W})/(2 A_W),\\ 
			t^W=t_1^W  \;\; or \;\; t^W=t_2^W, \\
			L'_W = Log[t^W]/2,\\
			F_W =( -2 (\sinh(2 K_1^W) \sinh(2 K_2^W) + \sinh(2 K^W) \sinh(2 L'_W))/
			\sinh(2 K_3^W))^{1/2}. \\
		\end{multline}
		This solution is valid along the trajectory

		\begin{equation}\label{Wu4_case1_1}
			\begin{gathered}
				\sinh( L_W)=(\exp(-K_1^W-K_2^W) \sinh(2 L'_W+K_W)-\exp(K_1^W+K_2^W) \sinh(K^W))/\\
				(2 \cosh(2 (K_1^W + K_2^W)) + 2 \cosh(2 (K^W + L'_W)))^{1/2}.
			\end{gathered} 	
		\end{equation}

		In this example

		\begin{equation}\label{Wu5_case1_1}
			\begin{gathered}
				K_1^W = K_{\omega^1,\omega^2} , \; \;  
				K_2^W =K_{\omega^0,\omega^2} ,  \; \;\\	
				K_3^W = K_{\omega^0,\omega^1},   \; \; 
				K^W =  K_{\omega^0,\omega^1,\omega^2}, ,
				t^W=t_1^W.
			\end{gathered} 	
		\end{equation}

		Then in the numerical example above for the "checkerboard-triangular" lattice we have
		
		\begin{equation}\label{Wu6_case1_1}
			\begin{gathered}
				F_W=\lambda_{max}=21.719383860264113, \\
				\sinh( L_W)=\sinh( 	K_{\omega^0}+	K_{\omega^1}+	K_{\omega^2})=36.27965089912661 .
			\end{gathered} 	
		\end{equation}

		I.e. in this numerical example, the parameters and resulting characteristics of this work coincide with the parameters and resulting characteristics calculated at the article \cite{Wu1985}. This also means indirect confirmation of the coincidence in the thermodynamic limit of the considered model and the model from  \cite{Wu1985}, where a periodic boundary condition is imposed in the horizontal direction only. 
		
		\begin{remark} \label{remark_chess}
			If we take a different meaning $W_1$, for example,  $W_1=3$, the parameters will change
			$	K_{\omega^0} = 0.5792929553213402, K_{\omega^1} = 2.505301112793842  $ 		
			, but the rest of the parameters will not change, including $\lambda_{max}$,  and the sum will not change
			$ 	K_{\omega^0}+	K_{\omega^1}= 3.084594068115182 ,$ that is, the total external field 
			$ 	K_{\omega^0}+	K_{\omega^1}+	K_{\omega^2}$  remains the same.
		\end{remark}

		 \subsubsection{ "Checkerboard-triangular" model with paired interactions }\label{checkerboard-triangular-paired}
		 	The Hamiltonian of the "checkerboard-triangular" model with paired interactions has the form
		 \begin{equation}\label{Chess_Hamiltonian_paired}
		 	\begin{gathered}
		 		\mathcal {H} (\sigma )= -\sum_{i=0,{\omega^{j}\subset \Omega_{\tau^i}}, j=0,1,2}^{L - 1} 		    	
		 		(	K_{\omega^{0} \omega^{1} } \sigma_{\omega^{0}} \sigma_{\omega^{1}}+
		 		K_{\omega^{0} \omega^{2} } \sigma_{\omega^{0}} \sigma_{\omega^{2}}+
		 		K_{\omega^{1} \omega^{2} } \sigma_{\omega^{1}} \sigma_{\omega^{2}}	).
		 	\end{gathered} 
		 \end{equation}
		   In this case from (\ref{K_even}) we have
		 
		 \begin{equation}\label{Chess_K_even}
		 	\begin{gathered}
		 		K_{\omega^{0} \omega^{1}}=-1/2 \;\; ((-1)  	A_{ (-1,+1) }),
		 	\end{gathered} 	
		 \end{equation}
		 where, taking into account (\ref{nuD_simplex_even_eigenvector}), we have
		 
		 \begin{equation}\label{Chess_A_even}
		 	\begin{gathered}
		 		A_{ (\sigma_{\omega^0},\sigma_{\omega^{1}}) }
		 		=
		 		\ln(			\sum_{\sigma_{\omega^{2}}} 		
		 		\exp(
		 		K_{\omega^{0} \omega^{2} } \sigma_{\omega^{0}} \sigma_{\omega^{2}}+
		 		K_{\omega^{1} \omega^{2} } \sigma_{\omega^{1}} \sigma_{\omega^{2}}	)
		 			 		)
		 		- \\
		 		\ln(			\sum_{\sigma_{\omega^{2}}} 		
		 		\exp(
		 		K_{\omega^{0} \omega^{2} }  \sigma_{\omega^{2}}+
		 		K_{\omega^{1} \omega^{2} }  \sigma_{\omega^{2}}	) 	)		 	
		 		,\\
		 		(\sigma_{\omega^0},\sigma_{\omega^{1}}) \in \{ (-1,+1),(+1,-1),(-1,-1)\}	.	
		 	\end{gathered} 	
		 \end{equation}
		 
		   This implies that
		   
		    \begin{equation}\label{K01_chess_even}
		   	\begin{gathered}
		   		K_{\omega^{0} \omega^{1} } = \frac{1}{2}
		   		\ln(\cosh(-K_{\omega^{0} \omega^{2} } + K_{\omega^{1} \omega^{2} })/\cosh(K_{\omega^{0} \omega^{2} } + K_{\omega^{1} \omega^{2} }))
		   		.	
		   	\end{gathered} 	
		   \end{equation}
		   
		   These solutions (\ref{K01_chess_even}) are also obtained in \cite{Wu1985}.
		 
		%%%%%%%%%%%%%%%%%%%%%%%%%%%%%%%%%%%%%%%%%%%%%%%%%%%%%%%%%%
		\subsection{Free energy  for 3D model  with the support of the Hamiltonian in the form of 3D simplex  }\label{3D_simplex}
		\hfill \break

		Let us consider the 3D simplex (triangular pyramid) as  $\Omega$ (Fig. 2):
	
		\begin{picture}(180,200) 		
			\put(20,55){\line(1,0){100}}
			\put(20,55){\line(0,1){100}}
			\put(20,55){\circle*{3}}
			\put(120,55){\circle*{3}}
			\put(0,40){${\tau^0}={\omega^0}$}
			\put(0,160){${\tau^{L_1 L_2}}={\omega^3}$}
			\put(100,40){${\tau^1}={{\omega^1}}$}		
			\put(20,55){\line(3,1){44}}		
			\put(120,55){\line(1,0){100}}
			\put(120,55){\line(-1,1){100}}			
			\put(63,69){\line(-1,2){43}}
			\put(63,69){\circle*{3}}
			\put(63,69){\line(4,-1){56}}			
			\put(50,73){${\tau^{L_1}}={\omega^2}$}
			\put(20,155){\circle*{3}}			
			\put(215,40){${\tau^2}$}
			\put(220,55){\circle*{3}}
			\put(55,10){Fig. 2}
		\end{picture}

		\begin{equation}\label{simplex_3D}
			\begin{gathered}
					\mathcal {S}_{3}=\{ \{{\omega}^0, \omega^1,\omega^2, \omega^3 \} , 
				{\omega}^0=\tau^0,  	{\omega}^1=\tau^1, 	{\omega}^2=\tau^{L_1},
				{\omega}^3=\tau^{L_1 L_2}=	{\omega}^{max}={\omega}^{r} \}.
			\end{gathered} 
		\end{equation}
		
		In this case $\Phi =\{{\omega}^0 \}.$

		 The Hamiltonian of the model has the form
		\begin{equation}\label{Hamiltonian_simplex_3D}
			\begin{gathered}
				\mathcal {H} (\sigma )= -\sum_{i=0,{\omega^{j}\subset \Omega_{\tau^i}}, j=0,1,2,3}^{L-1} 		    	
				(	K_{\omega^{0}} 	\sigma_{\omega^{0}}+
				K_{\omega^{1}} 	\sigma_{\omega^{1}}+	K_{\omega^{2}} 	\sigma_{\omega^{2}}+	K_{\omega^{3}} 	\sigma_{\omega^{3}}+\\ 
				K_{\omega^{0} \omega^{1} } \sigma_{\omega^{0}} \sigma_{\omega^{1}}+
				K_{\omega^{0} \omega^{2} } \sigma_{\omega^{0}} \sigma_{\omega^{2}}+
				K_{\omega^{0} \omega^{3} } \sigma_{\omega^{0}} \sigma_{\omega^{3}}+			K_{\omega^{1} \omega^{2} } \sigma_{\omega^{1}} \sigma_{\omega^{2}}+ \\
				K_{\omega^{1} \omega^{3} } \sigma_{\omega^{1}} \sigma_{\omega^{3}}+		
				K_{\omega^{2} \omega^{3} } \sigma_{\omega^{2}} \sigma_{\omega^{3}}+	 
				K_{\omega^{0} \omega^{1} \omega^{2} } \sigma_{\omega^{0}}  \sigma_{\omega^{1}} \sigma_{\omega^{2}}	+
				K_{\omega^{0} \omega^{1} \omega^{3} } \sigma_{\omega^{0}}  \sigma_{\omega^{1}} \sigma_{\omega^{3}}+ \\
				K_{\omega^{0} \omega^{2} \omega^{3} } \sigma_{\omega^{0}}  \sigma_{\omega^{2}} \sigma_{\omega^{3}}+ 
				K_{\omega^{1} \omega^{2} \omega^{3} } \sigma_{\omega^{1}}  \sigma_{\omega^{2}} \sigma_{\omega^{3}}+
				K_{\omega^{0} \omega^{1} \omega^{2}  \omega^{3}} \sigma_{\omega^{0}}  \sigma_{\omega^{1}} \sigma_{\omega^{2}} \sigma_{\omega^{3}}),
			\end{gathered} 
		\end{equation}
		wherein $ L=L_1 L_2 L_3$, and this is the 3D model .
		As $\Phi =\{{\omega}^0 \},$ the eigenvector has the form

		\begin{equation}\label{3D_eigenvector_1_simplex}
			\begin{gathered}
				w_{ (\sigma_{\omega^0},\sigma_{\omega^{1}},\sigma_{\omega^{2}})  }(\sigma_{\omega^{0}}=\sigma_{\phi^{1}}) =\\
				(W_0 (\sigma_{\omega^{0}} + 1)/2 + W_1 (1 - \sigma_{\omega^{0}})/2) 
				\\					    			
			\end{gathered} 	
		\end{equation}
	
			Let us write the system of 8 equations (\ref{main_system_small}) for this case.
	
	\begin{equation}\label{main_system_small_simplex_3D}
		\begin{gathered} 
			{\lambda_{max}}	w_{ (\sigma_{\omega^0},\sigma_{\omega^{1}},\sigma_{\omega^{2}})  } =
			\sum_{\sigma_{\omega^{3}}} 
			\Theta_{\{ (\sigma_{\omega^0},\sigma_{\omega^{1}},\sigma_{\omega^{2}}),
				(\sigma_{\omega^1},\sigma_{\omega^{2}},\sigma_{\omega^{3}})  \} }
			w_{ (\sigma_{\omega^1},\sigma_{\omega^{2}},\sigma_{\omega^{3}})}
			,		\\
			\sigma_{\omega^j} \in X=\{+1,-1 \}, j=0,1,2,3.	
		\end{gathered} 	
	\end{equation}

	Then from  (\ref{K}), (\ref{A_right}), (\ref{K_nuD_simplex}), (\ref{A_nuD_simplex}) we have
	
	\begin{equation}\label{K_simplex_3D}
		\begin{gathered}
			K_{\omega^{0}}=-1/8 \;\; ((-1) 	A_{ (-1,+1,+1) } +(+1) 	A_{ (+1,-1,+1) } + 
			(-1) 	A_{ (-1,-1,+1) }+ \\
			(+1) 	A_{ (+1,+1,-1) }+
			(-1) 	A_{ (-1,+1,-1) } +(+1) 	A_{ (+1,-1,-1) } + 
			(-1) 	A_{ (-1,-1,-1) }
			) , \\
			K_{\omega^{1}}=-1/8 \;\; ((+1) 	A_{ (-1,+1,+1) } +(-1) 	A_{ (+1,-1,+1) } + 
			(-1) 	A_{ (-1,-1,+1) }+ \\
			(+1) 	A_{ (+1,+1,-1) }+
			(+1) 	A_{ (-1,+1,-1) } +(-1) 	A_{ (+1,-1,-1) } + 
			(-1) 	A_{ (-1,-1,-1) }
			) , \\
			K_{\omega^{2}}=-1/8 \;\; ((+1) 	A_{ (-1,+1,+1) } +(+1) 	A_{ (+1,-1,+1) } + 
			(+1) 	A_{ (-1,-1,+1) }+ \\
			(-1) 	A_{ (+1,+1,-1) }+
			(-1) 	A_{ (-1,+1,-1) } +(-1) 	A_{ (+1,-1,-1) } + 
			(-1) 	A_{ (-1,-1,-1) }
			) , \\
			K_{\omega^{0} \omega^{1}}=-1/8 \;\; ((-1) (+1) 	A_{ (-1,+1,+1) } +(+1) (-1) 	A_{ (+1,-1,+1) } + \\
			(-1) (-1)	A_{ (-1,-1,+1) }+ 
			(+1) (+1)	A_{ (+1,+1,-1) }+\\
			(-1) (+1)	A_{ (-1,+1,-1) } +(+1) (-1) 	A_{ (+1,-1,-1) } + 
			(-1) (-1)	A_{ (-1,-1,-1) }
			) , \\
			K_{\omega^{0} \omega^{2}}=-1/8 \;\; ((-1) (+1) 	A_{ (-1,+1,+1) } +(+1) (+1) 	A_{ (+1,-1,+1) } + \\
			(-1) (+1)	A_{ (-1,-1,+1) }+ 
			(+1) (-1)	A_{ (+1,+1,-1) }+\\
			(-1) (-1)	A_{ (-1,+1,-1) } +(+1) (-1) 	A_{ (+1,-1,-1) } + 
			(-1) (-1)	A_{ (-1,-1,-1) }
			)  ,\\
			K_{\omega^{1} \omega^{2}}=-1/8 \;\; ((+1) (+1) 	A_{ (-1,+1,+1) } +(-1) (+1) 	A_{ (+1,-1,+1) } + \\
			(-1) (+1)	A_{ (-1,-1,+1) }+ 
			(+1) (-1)	A_{ (+1,+1,-1) }+\\
			(+1) (-1)	A_{ (-1,+1,-1) } +(-1) (-1) 	A_{ (+1,-1,-1) } + 
			(-1) (-1)	A_{ (-1,-1,-1) }
			) , \\
			K_{\omega^{0} \omega^{1} \omega^{2}}=-1/8 \;\; ((-1) (+1) (+1)	A_{ (-1,+1,+1) } +(+1) (-1) (+1)	A_{ (+1,-1,+1) } + \\
			(-1) (-1) (+1)	A_{ (-1,-1,+1) }+ 
			(+1) (+1) (-1)	A_{ (+1,+1,-1) }+\\
			(-1) (+1) (-1)	A_{ (-1,+1,-1) } + 
			(+1) (-1) (-1)	A_{ (+1,-1,-1) } + \\
			(-1) (-1) (-1)	A_{ (-1,-1,-1) }
			) , \\
		\end{gathered} 	
	\end{equation}
	where

	\begin{equation}\label{A_right_simplex_3D}
		\begin{gathered}
			A_{ (\sigma_{\omega^0},\sigma_{\omega^{1}},\sigma_{\omega^{2}}) }
			=
			\ln(			\sum_{\sigma_{\omega^{3}}} 		
			\exp	(		K_{\omega^{3}} 	\sigma_{\omega^{3}}+ 
			K_{\omega^{0} \omega^{3} } \sigma_{\omega^{0}} \sigma_{\omega^{3}}+			
			K_{\omega^{1} \omega^{3} } \sigma_{\omega^{1}} \sigma_{\omega^{3}}+	\\	
			K_{\omega^{2} \omega^{3} } \sigma_{\omega^{2}} \sigma_{\omega^{3}}+	 
			K_{\omega^{0} \omega^{1} \omega^{3} } \sigma_{\omega^{0}}  \sigma_{\omega^{1}} \sigma_{\omega^{3}}+ 
			K_{\omega^{0} \omega^{2} \omega^{3} } \sigma_{\omega^{0}}  \sigma_{\omega^{2}} \sigma_{\omega^{3}}+ \\
			K_{\omega^{1} \omega^{2} \omega^{3} } \sigma_{\omega^{1}}  \sigma_{\omega^{2}} \sigma_{\omega^{3}}+
			K_{\omega^{0} \omega^{1} \omega^{2}  \omega^{3}} \sigma_{\omega^{0}}  \sigma_{\omega^{1}} \sigma_{\omega^{2}} \sigma_{\omega^{3}}) 
			w_{ (\sigma_{\omega^1},\sigma_{\omega^{2}},\sigma_{\omega^{3}})})
			- \\
			\ln(		\sum_{\sigma_{\omega^{3}}} 	
			\exp	(
			K_{\omega^{3}} 	\sigma_{\omega^{3}}+ 
			K_{\omega^{0} \omega^{3} } \sigma_{\omega^{3}}+			
			K_{\omega^{1} \omega^{3} }  \sigma_{\omega^{3}}+		
			K_{\omega^{2} \omega^{3} }  \sigma_{\omega^{3}}+	 
			K_{\omega^{0} \omega^{1} \omega^{3} }  \sigma_{\omega^{3}}+ \\
			K_{\omega^{0} \omega^{2} \omega^{3} }  \sigma_{\omega^{3}}+ 
			K_{\omega^{1} \omega^{2} \omega^{3} } \sigma_{\omega^{3}}+
			K_{\omega^{0} \omega^{1} \omega^{2}  \omega^{3}} \sigma_{\omega^{3}}) 
			w_{ (\sigma_{\omega^1}=+1,\sigma_{\omega^{2}}=+1,\sigma_{\omega^{3}})}
			)		 		
			+\\
			\ln(	w_{ (\sigma_{\omega^0}=+1,\sigma_{\omega^{1}}=+1 ,\sigma_{\omega^{2}}=+1) })- 	
			\ln(w_{ (\sigma_{\omega^0},\sigma_{\omega^{1}},\sigma_{\omega^{2}})  } )
			,\\
			(\sigma_{\omega^0},\sigma_{\omega^{1}},\sigma_{\omega^{2}}) \neq  (+1,+1,+1).		
		\end{gathered} 	
	\end{equation}	
			
		 For Hamiltonians, containing interactions of only products of an even number of spins, we have		  
		  
		  	\begin{equation}\label{K_simplex_3D_even}
		  	\begin{gathered}
		  		K_{\omega^{0} \omega^{1}}=-1/4 \;\; ((-1) (+1) 	A_{ (-1,+1,+1) } +\\
		  		(+1) (-1) 	A_{ (+1,-1,+1) } + 
		  		(-1) (-1)	A_{ (-1,-1,+1) }), \\
		  		K_{\omega^{0} \omega^{2}}=-1/4 \;\; ((-1) (+1) 	A_{ (-1,+1,+1) } +\\
		  		(+1) (+1) 	A_{ (+1,-1,+1) } + 
		  		(-1) (+1)	A_{ (-1,-1,+1) }) ,\\
		  		K_{\omega^{1} \omega^{2}}=-1/4 \;\; ((+1) (+1) 	A_{ (-1,+1,+1) } +\\
		  		(-1) (+1) 	A_{ (+1,-1,+1) } + 
		  		(-1) (+1)	A_{ (-1,-1,+1) }) , \\
		  		K_{\omega^{0} \omega^{1} \omega^{2}}=0 , 
		  			K_{\omega^{0}}=0, 
		  		K_{\omega^{1}}=0 ,
		  		K_{\omega^{2}}=0 , 
		  	\end{gathered} 	
		  \end{equation}
		  where
		  		  
		  \begin{equation}\label{A_right_simplex_3D_even}
		  	\begin{gathered}
		  		A_{ (\sigma_{\omega^0},\sigma_{\omega^{1}},\sigma_{\omega^{2}}) }
		  		=
		  		\ln(			\sum_{\sigma_{\omega^{3}}} 		
		  		\exp	(	
		  		K_{\omega^{0} \omega^{3} } \sigma_{\omega^{0}} \sigma_{\omega^{3}}+			
		  		K_{\omega^{1} \omega^{3} } \sigma_{\omega^{1}} \sigma_{\omega^{3}}+\\		
		  		K_{\omega^{2} \omega^{3} } \sigma_{\omega^{2}} \sigma_{\omega^{3}}+	 
		  		K_{\omega^{0} \omega^{1} \omega^{2}  \omega^{3}} \sigma_{\omega^{0}}  \sigma_{\omega^{1}} \sigma_{\omega^{2}} \sigma_{\omega^{3}}))
		  		- \\
		  		\ln(		\sum_{\sigma_{\omega^{3}}} 	
		  		\exp	(
		  				  		K_{\omega^{0} \omega^{3} } \sigma_{\omega^{3}}+			
		  		K_{\omega^{1} \omega^{3} }  \sigma_{\omega^{3}}+		
		  		K_{\omega^{2} \omega^{3} }  \sigma_{\omega^{3}}+	 
		  			  			  		K_{\omega^{0} \omega^{1} \omega^{2}  \omega^{3}} \sigma_{\omega^{3}}) ).		
		  	\end{gathered} 	
		  \end{equation}

		 %%%%%%%%%%%%%%%%%%%%%%%%%%%%%%%%%%%%%%%%%%%%%%%%%%%%%%%%%%%%%%%%%%%%%%%%%%%%%%%%%
		\section{Free energy  for $\nu$ - dimensional models  with the support of the Hamiltonian in the form of $\nu$ - dimensional  cube }\label{nuD}
			
	Let us consider the unit $\nu$ - dimensional cube as $\Omega$

	\begin{equation}\label{cube_nu}
		\begin{gathered}			
			\Pi_{\nu}=\{t=(t_1,t_2,...,t_{\nu}) \in\mathcal {L}_{\nu} : t_i=0,1, i=1,2,...,\nu\}=\\		
			=\{{\omega}^0, \omega^1,...,\omega^{2^{\nu}-1} \} , \\
			{\omega}^0=\tau^0,  	{\omega}^1=\tau^1, 	{\omega}^2=\tau^{L_1},
			{\omega}^3=\tau^{L_1+1}, 	{\omega}^4=\tau^{L_1 L_2}, \\
			{\omega}^5=\tau^{L_1 L_2 +1}, 	{\omega}^6=\tau^{L_1 L_2+L_1},	
			{\omega}^7=\tau^{L_1 L_2+L_1+1},..., \\
			{\omega}^{2^{\nu}-1}=\tau^{L_1 L_2...L_{\nu -1}+L_1 L_2 ... L_{\nu -2}+...+L_1 L_2+L_1+1}=	{\omega}^{max}={\omega}^{r}.
		\end{gathered} 
	\end{equation}
			
			Then
				
		\begin{equation}\label{phi_cube_nu}
			\begin{gathered}
				\Phi =\{{\omega}^0, \omega^2, \omega^4, \omega^6,...,\omega^{2^{\nu}-2}\} .
			\end{gathered} 
		\end{equation}

		The Hamiltonian of the model has the form  (\ref{Hamiltonian1}) ,		 
		where $\tau^i=(\tau_1^i,\tau_2^i,...,\tau_{\nu}^i) \in  \mathcal {L}_{\nu} .   $ 
		
		From (\ref{phi_cube_nu}) eigenvector components  $	\overrightarrow{w}$  have the form

			\begin{equation}\label{w_cube_nu}
			\begin{gathered} 
				w_{ (\sigma_{\omega^0},\sigma_{\omega^{1}},...,\sigma_{\omega^{{2^{\nu}}-2}})  }(\sigma_{\omega^{2 k - 2}}=\sigma_{\phi^{k}}, k=1, 2, ..., {2^{\nu-1}}) =
				W_{ (\sigma_{\phi^1},\sigma_{\phi^{2}},...,\sigma_{\phi^{{2^{\nu -1}}}})}.\\		     			     			
			\end{gathered} 	
		\end{equation}

		It is convenient to take $W_0=1$ as the normalization of the eigenvector $	\overrightarrow{w}$.
		
	The solution  (\ref{K}) for the case of $\nu$-dimensional cube $\Omega=	\Pi_{\nu}$ can be written in the form

		\begin{equation}\label{K_nuD}
			\begin{gathered}			
				K_{\omega^{j_1} , \omega^{j_2} , ..., \omega^{j_s}}=\\
				-(\sum_{\{\sigma_{\omega^0},\sigma_{\omega^1},...,\sigma_{\omega^{2^{\nu}-2}}\}}
				\sigma_{\omega^{j_1}}\sigma_{\omega^{j_2}}...\sigma_{\omega^{j_s}}	
				A_{\sigma_{\omega^0} ,\sigma_{\omega^1},...,\sigma_{{\omega^{2^{\nu}-2}}} })/2^{(2^{\nu}-1)},\\
				\{\sigma_{\omega^0} ,\sigma_{\omega^1},...,\sigma_{\omega^{2^{\nu}-2}} \} 	\neq
				(+1,+1,...,+1 )	, 			
			\end{gathered} 	
		\end{equation}		
		
		where 		
		
		\begin{equation}\label{A_nuD}
			\begin{gathered}
				A_{\sigma_{\omega^0} ,\sigma_{\omega^1},...,\sigma_{\omega^{2^{\nu}-2}} }=\\
				\ln(			\sum_{\sigma_{\omega^{2^{\nu}-1}}} 		
				\exp(  \sum_{\{\omega^{j_1},\omega^{j_2},...,\omega^{2^{\nu}-1}\}\subset \Omega_{\tau^0}}
				K_{\omega^{j_1} , \omega^{j_2} , ..., \omega^{2^{\nu}-1}}
				\sigma_{\omega^{j_1}}\sigma_{\omega^{j_2}}...
				\sigma_{\omega^{2^{\nu}-1}}) \cdot \\		
				w_{ (\sigma_{\omega^{1}},\sigma_{\omega^{2}},...
				,
					\sigma_{\omega^{2^{\nu}-2}},\sigma_{\omega^{2^{\nu}-1}})  }
				)
				- \\
				\ln(			\sum_{\sigma_{\omega^{2^{\nu}-1}}} 		
				\exp(  \sum_{\{\omega^{j_1},\omega^{j_2},...,\omega^{2^{\nu}-1}\}\subset \Omega_{\tau^0}}
				K_{\omega^{j_1} , \omega^{j_2} , ..., \omega^{2^{\nu}-1}}\sigma_{\omega^{2^{\nu}-1}}) \cdot \\
				w_{ (\sigma_{\omega^1}=+1,\sigma_{\omega^{2}}=+1,...,,\sigma_{\omega^{2^{\nu}-2}}=+1,\sigma_{\omega^{2^{\nu}-1}})})  +\\
				\ln(	w_{ (\sigma_{\omega^0}=+1,\sigma_{\omega^{1}}=+1,...,\sigma_{\omega^{2^{\nu}-2}}=+1)  })- 	
				\ln(w_{ (\sigma_{\omega^0},\sigma_{\omega^{1}},\sigma_{\omega^{2}},...,	\sigma_{\omega^{2^{\nu}-2}})  }  ).
			\end{gathered} 	
		\end{equation}	 
	
	By revising $\nu$ - dimensional models  with the support of the Hamiltonian in the form of $\nu$ - dimensional cube, containing nonzero only the interaction coefficients of the product of an even number of spins, it is necessary to take free variables in (\ref{A_nuD}), free variables should contain ${ \omega^{2^{\nu}-1}}$, and corresponding to the interaction of an odd number of spins equal to zero:
	$	K_{\omega^{j_1} , \omega^{j_2} , ..., \omega^{j_{2 p}} , \omega^{2^{\nu}-1}}=0 $ . The eigenvector $	\overrightarrow{w}$ should look like (\ref{w_cube_nu}) and satisfy the central symmetry condition  (\ref{w_even})  . Then from the case (\ref{even}) in the formula (\ref{K_nuD}) the interaction coefficients of the product of an odd number of spins are equal to zero:	$	K_{\omega^{j_1} , \omega^{j_2} , ..., \omega^{j_{2 p-1}}}=0 $ for ${\{\omega^{j_1},\omega^{j_2},...,\omega^{j_{2 p -1}}\}\subset \Omega'_{\tau^0}}. $ 
	We can  simplify the formula (\ref{K}) for this case:

	\begin{equation}\label{K_even_cube} 
		\begin{gathered}			
			K_{\omega^{j_1} , \omega^{j_2} , ..., \omega^{j_{2 p}}}=\\
			-(\sum_{\{\sigma_{\omega^0},\sigma_{\omega^1},...,\sigma_{\omega^{2^{\nu}-3}}\}}
			\sigma_{\omega^{j_1}}\sigma_{\omega^{j_2}}...\sigma_{\omega^{j_{2 p}}}	
			A_{\sigma_{\omega^0} ,\sigma_{\omega^1},...,\sigma_{\omega^{2^{\nu}-3}},
				\sigma_{\omega^{2^{\nu}-2}}=+1 })/2^{2^{\nu}-2},\\
			\{\sigma_{\omega^0} ,\sigma_{\omega^1},...,\sigma_{\omega^{2^{\nu}-2}} \} 	\neq
			(+1,+1,...,+1 )	,
			{\{\omega^{j_1},\omega^{j_2},...,\omega^{j_{2 p}}\}\subset \Omega'_{\tau^0}},\\ 	
			K_{\omega^{j_1} , \omega^{j_2} , ..., \omega^{j_{2 p-1}}}=	0,\\
			{\{\omega^{j_1},\omega^{j_2},...,\omega^{j_{2 p}}\}\subset \Omega'_{\tau^0}},	  
				{\{\omega^{j_1},\omega^{j_2},...,\omega^{j_{2 p -1}}\}\subset \Omega'_{\tau^0}},	 				
		\end{gathered} 	
	\end{equation}
	
	where   $	A_{\sigma_{\omega^0} ,\sigma_{\omega^1},...,\sigma_{\omega^{2^{\nu}-3}},
		\sigma_{\omega^{2^{\nu}-2}}=+1 } $ we find by the formulas (\ref{A_nuD}) with a centrally symmetric vector $	\overrightarrow{ w}$ (\ref{w_even}).

	 Then let us consider as examples the lattice models for $\nu=2$ and $\nu=3.$ 
%%%%%%%%%%%%%%%%%%%%%%%%%%%%%%%%%%%%%%%%%%%%%%%%%%%%%%%%%%%%%%%%%%%%%%%%
		\subsection{Free energy  for cube  model on the 2D lattice. General case }\label{2D}
			
		Let us consider the lattice (\ref{lattice_1}) - (\ref{lattice_3}) for  $\nu=2$ (Fig.2).  
	
		\begin{picture}(180,180) 		
			\put(20,55){\line(1,0){100}}
			\put(20,55){\line(0,1){100}}
			\put(20,55){\circle*{3}}
			\put(120,55){\circle*{3}}
			\put(0,40){${\tau^0}={\omega^0}$}
			\put(100,40){${\tau^1}={{\omega^1}}$}	
			\put(215,40){${\tau^2}$}
			\put(20,155){\line(1,0){100}}
			\put(220,55){\circle*{3}}	
			\put(120,55){\line(0,1){100}}	
			\put(120,55){\line(1,0){100}}		
			\put(20,155){\circle*{3}}
			\put(120,155){\circle*{3}}
			\put(0,165){${\tau^{L_1}}={\omega^2}$}
			\put(110,165){${\tau^{L_1 +1}}={\omega^3}$}		
			\qbezier[460](220,55)(520,90)(20,155)
			\put(55,10){Fig. 2}
		\end{picture}
			
		The Hamiltonian of the model has the form 
		\begin{equation}\label{2D_Hamiltonian1}
			\begin{gathered}
				\mathcal {H} (\sigma )= -\sum_{i=0,{\omega^{j}\subset \Omega_{\tau^i}}, j=0,1,2,3}^{L-1} 		    	
				(	K_{\omega^{0}} 	\sigma_{\omega^{0}}+
				K_{\omega^{1}} 	\sigma_{\omega^{1}}+	K_{\omega^{2}} 	\sigma_{\omega^{2}}+	K_{\omega^{3}} 	\sigma_{\omega^{3}}+\\ 
				K_{\omega^{0} \omega^{1} } \sigma_{\omega^{0}} \sigma_{\omega^{1}}+
				K_{\omega^{0} \omega^{2} } \sigma_{\omega^{0}} \sigma_{\omega^{2}}+
				K_{\omega^{0} \omega^{3} } \sigma_{\omega^{0}} \sigma_{\omega^{3}}+			K_{\omega^{1} \omega^{2} } \sigma_{\omega^{1}} \sigma_{\omega^{2}}+ \\
				K_{\omega^{1} \omega^{3} } \sigma_{\omega^{1}} \sigma_{\omega^{3}}+		
				K_{\omega^{2} \omega^{3} } \sigma_{\omega^{2}} \sigma_{\omega^{3}}+	 
				K_{\omega^{0} \omega^{1} \omega^{2} } \sigma_{\omega^{0}}  \sigma_{\omega^{1}} \sigma_{\omega^{2}}	+
				K_{\omega^{0} \omega^{1} \omega^{3} } \sigma_{\omega^{0}}  \sigma_{\omega^{1}} \sigma_{\omega^{3}}+ \\
				K_{\omega^{0} \omega^{2} \omega^{3} } \sigma_{\omega^{0}}  \sigma_{\omega^{2}} \sigma_{\omega^{3}}+ 
				K_{\omega^{1} \omega^{2} \omega^{3} } \sigma_{\omega^{1}}  \sigma_{\omega^{2}} \sigma_{\omega^{3}}+
				K_{\omega^{0} \omega^{1} \omega^{2}  \omega^{3}} \sigma_{\omega^{0}}  \sigma_{\omega^{1}} \sigma_{\omega^{2}} \sigma_{\omega^{3}}),
			\end{gathered} 
		\end{equation}
		
		where $ L=L_1 L_2 .$

		From the formulas (\ref{Omega}) , (\ref{Phi}) we have $\Omega= \{ \omega^0,  \omega^1,  \omega^2 ,   \omega^3 \},$ $r=2$,
		$\Phi=\{\phi^1= \omega^0 , \phi^2= \omega^2\}$.  Consequently,
		\begin{equation}\label{2D_eigenvector_1}
			\begin{gathered}
				w_{ (\sigma_{\omega^0},\sigma_{\omega^{1}},\sigma_{\omega^{2}})  }(\sigma_{\omega^{0}}=\sigma_{\phi^{1}}, \sigma_{\omega^{2}}=\sigma_{\phi^{2}}) =\\
				(W_0 (\sigma_{\omega^{0}} + 1)/2 + W_1 (1 - \sigma_{\omega^{0}})/2) 
				(1 + \sigma_{\omega^{2}})/2 + \\
				(W_3 (\sigma_{\omega^{0}} + 1)/2 + W_4 (1 - \sigma_{\omega^{0}})/2) 
				(1 - \sigma_{\omega^{2}})/2			 . \\					    			
			\end{gathered} 	
		\end{equation}

		It means that the eigenvector of the elementary transfer matrix $ \Theta=\Theta_{p,q} $, corresponding to
		the largest eigenvalue $\lambda_{max} $, we represent in the following form  \\
		 
		\begin{equation}\label{2D_eigenvector_2}   
			\overrightarrow{V}=
			(\underbrace{W_0,W_1,\dots,W_0,W_1 }_{2^{L_1}},
			\underbrace{W_3,W_4,\dots,W_3,W_4 }_{2^{L_1}})^T
			.
		\end{equation}	
				  
		The system of 8 equations (\ref{main_system_small}) for this case coincide with  (\ref{main_system_small_simplex_3D}), the formulas for finding coefficients $K_{\omega^{j_1} , \omega^{j_2} , ..., \omega^{j_s}}$ and the right side also coincide with (\ref{K_simplex_3D}) ,
		(\ref{A_right_simplex_3D}).  Only vectors $\overrightarrow{w}$ will differ, for the case under consideration, we use the formulas (\ref{2D_eigenvector_1}), (\ref{2D_eigenvector_2}). As a special case, solutions with an eigenvector (\ref{3D_eigenvector_1_simplex}) for a three-dimensional simplex are also suitable at this point. But not vice versa.
	
		%%%%%%%%%%%%%%%%%%%%%%%%%%%
			
		\textbf{Example.} {2D Cube model}
			
		We define the free parameters:
		$  K_{\omega^0, \omega^1, \omega^{2}, \omega^{3}}= 1.0123, 
		K_{\omega^0, \omega^1,  \omega^{3}}=  -0.013 ,$
		$
		K_{\omega^0,  \omega^{2}, \omega^{3}}=1.5,
		K_{\omega^0,  \omega^{3}}=0.693,
		K_{ \omega^1, \omega^{2}, \omega^{3}}=	0.64,		 
		K_{ \omega^1, \omega^{3}}=0.13,$
		$
		K_{ \omega^{2}, \omega^{3}}=0.23,
		K_{ \omega^{3}}=1.3, 
		W_0=1,
		W_1=9,
		W_2=0.03,		 
		W_3=2.4	. $		
		
		Other parameters are calculated by the formulas (\ref{K_simplex_3D})-(\ref{A_right_simplex_3D})  :
		\begin{equation}\label{2D_calculate}
			\begin{gathered}
				K_{\omega^0}=  -2.2956475107799923,\;\; K_{\omega^1}= 0.9727655738251864, \\ 
				K_{\omega^2}= 0.9379786258612406, \;\;
				{ K_{\omega^0, \omega^{1}}=-0.06890531130394228 }, \\
				{ K_{\omega^0, \omega^{2}}= -0.8333056330589352},\;\;
				{ K_{\omega^1, \omega^{2}}= -0.5610350805896709},		\\ 				
				K_{\omega^0, \omega^{1},  \omega^{2}}=-1.010841276086467,\;\;
				\lambda_{max}= 13.919766727642834 , \\
				- f/ {(kT)}= 2.6333098966942203 ,
			\end{gathered} 	
		\end{equation}
		where $f$ is free energy per one lattice site.
				
		\begin{remark} \label{remark_2D}
			If we take in (\ref{A_right_simplex_3D})
		\end{remark}
		\begin{equation}\label{2D_K_0}
			\begin{gathered}
				K_{\omega^{0}  \omega^{3}}=0, 
				K_{\omega^{0} \omega^{1}  \omega^{3}}=0, 	K_{\omega^{0} \omega^{2}  \omega^{3}}=0, 	K_{\omega^{0} \omega^{1} \omega^{2}  \omega^{3}}=0, 		
			\end{gathered} 	
		\end{equation}
		we get  solutions (\ref{K_simplex_3D}) on a complete triangular lattice. If we add the condition $K_{ \omega^{1} \omega^{2}  \omega^{3}}=0$ to the condition (\ref{2D_K_0}), we get the solution on a chess-triangular lattice.

			%%%%%%%%%%%%%%%%%%%%%
				\subsection{Free energy  for cube  models invariant under the reversal of signs of all spins on the 2D lattice }\label{2D_even}

			Now let the Hamiltonian (\ref{2D_Hamiltonian1})
	has nonzero coefficients only for products of an even number of terms
	\begin{equation}\label{2D_Hamiltonian1_even}
	\begin{gathered}
		\mathcal {H} (\sigma )= -\sum_{i=0,{\omega^{j}\subset \Omega_{\tau^i}}, j=0,1,2,3}^{L-1} 		    	
		(	K_{\omega^{0} \omega^{1} } \sigma_{\omega^{0}} \sigma_{\omega^{1}}+
		K_{\omega^{0} \omega^{2} } \sigma_{\omega^{0}} \sigma_{\omega^{2}}+
		K_{\omega^{0} \omega^{3} } \sigma_{\omega^{0}} \sigma_{\omega^{3}}+	\\		K_{\omega^{1} \omega^{2} } \sigma_{\omega^{1}} \sigma_{\omega^{2}}+ 
		K_{\omega^{1} \omega^{3} } \sigma_{\omega^{1}} \sigma_{\omega^{3}}+		
		K_{\omega^{2} \omega^{3} } \sigma_{\omega^{2}} \sigma_{\omega^{3}}+	 
			K_{\omega^{0} \omega^{1} \omega^{2}  \omega^{3}} \sigma_{\omega^{0}}  \sigma_{\omega^{1}} \sigma_{\omega^{2}} \sigma_{\omega^{3}}).
	\end{gathered} 
\end{equation}

Then from  (\ref{w_even}) we have
	\begin{equation}\label{2D_eigenvector_1_even}
	\begin{gathered}
		w_{ (\sigma_{\omega^0},\sigma_{\omega^{1}},\sigma_{\omega^{2}})  }(\sigma_{\omega^{0}}=\sigma_{\phi^{1}}, \sigma_{\omega^{2}}=\sigma_{\phi^{2}}) =\\
		(W_0 (\sigma_{\omega^{0}} + 1)/2 + W_1 (1 - \sigma_{\omega^{0}})/2) 
		(1 + \sigma_{\omega^{2}})/2 + \\
		(W_1 (\sigma_{\omega^{0}} + 1)/2 + W_0 (1 - \sigma_{\omega^{0}})/2) 
		(1 - \sigma_{\omega^{2}})/2			 . \\					    			
	\end{gathered} 	
\end{equation}		
		From   (\ref{K}), (\ref{A_right}), (\ref{K_nuD_simplex}), (\ref{A_nuD_simplex}) we have
		
		\begin{equation}\label{K_2D_even}
			\begin{gathered}			
				K_{\omega^{0} \omega^{1}}=-1/4 \;\; ((-1) (+1) 	A_{ (-1,+1,+1) } +\\
				(+1) (-1) 	A_{ (+1,-1,+1) } + 
				(-1) (-1)	A_{ (-1,-1,+1) }) , \\
				K_{\omega^{0} \omega^{2}}=-1/4 \;\; ((-1) (+1) 	A_{ (-1,+1,+1) } +\\(+1) (+1) 	A_{ (+1,-1,+1) } + 
				(-1) (+1)	A_{ (-1,-1,+1) })  ,\\
				K_{\omega^{1} \omega^{2}}=-1/4 \;\; ((+1) (+1) 	A_{ (-1,+1,+1) } +\\
				(-1) (+1) 	A_{ (+1,-1,+1) } + 
				(-1) (+1)	A_{ (-1,-1,+1) }) , \\
						\end{gathered} 	
		\end{equation}
		where

		\begin{equation}\label{A_right_2D_even}
			\begin{gathered}
				A_{ (\sigma_{\omega^0},\sigma_{\omega^{1}},\sigma_{\omega^{2}}) }
				=
				\ln(			\sum_{\sigma_{\omega^{3}}} 		
				\exp	(	
				K_{\omega^{0} \omega^{3} } \sigma_{\omega^{0}} \sigma_{\omega^{3}}+			
				K_{\omega^{1} \omega^{3} } \sigma_{\omega^{1}} \sigma_{\omega^{3}}+	\\	
				K_{\omega^{2} \omega^{3} } \sigma_{\omega^{2}} \sigma_{\omega^{3}}+ 
						K_{\omega^{0} \omega^{1} \omega^{2}  \omega^{3}} \sigma_{\omega^{0}}  \sigma_{\omega^{1}} \sigma_{\omega^{2}} \sigma_{\omega^{3}}) 
				w_{ (\sigma_{\omega^1},\sigma_{\omega^{2}},\sigma_{\omega^{3}})})
				- \\
				\ln(		\sum_{\sigma_{\omega^{3}}} 	
				\exp	(
				K_{\omega^{0} \omega^{3} } \sigma_{\omega^{3}}+			
				K_{\omega^{1} \omega^{3} }  \sigma_{\omega^{3}}+		
				K_{\omega^{2} \omega^{3} }  \sigma_{\omega^{3}}+	\\ 
				K_{\omega^{0} \omega^{1} \omega^{2}  \omega^{3}} \sigma_{\omega^{3}}) 
				w_{ (\sigma_{\omega^1}=+1,\sigma_{\omega^{2}}=+1,\sigma_{\omega^{3}})}
				)		 		
				+\\
				\ln(	w_{ (\sigma_{\omega^0}=+1,\sigma_{\omega^{1}}=+1 ,\sigma_{\omega^{2}}=+1) })- 	
				\ln(w_{ (\sigma_{\omega^0},\sigma_{\omega^{1}},\sigma_{\omega^{2}})  } )
				,\\
				(\sigma_{\omega^0},\sigma_{\omega^{1}},\sigma_{\omega^{2}}) \neq  (+1,+1,+1).		
			\end{gathered} 	
		\end{equation}

		\begin{remark} \label{remark_2D_even}
		If we take in (\ref{A_right_2D_even})
		\end{remark}
		\begin{equation}\label{2D_K_0_even}
			\begin{gathered}
				K_{\omega^{0}  \omega^{3}}=0, 
			K_{\omega^{0} \omega^{1} \omega^{2}  \omega^{3}}=0, 		
				\end{gathered} 	
		\end{equation}
		we obtain disordered parametric solutions on a triangular lattice for the Hamiltonians invariant under the reversal of signs of all spins.  
	%%%%%%%%%%%%%%%%%%%%%%%%%%%%%%%%%%%%%%%%%%%%%%%%%%%%%%%%%%%%%
		\subsection{Free energy  for cube model on the 3D lattice. General case }\label{3D}
	
			\hfill \break
	
		Let us consider the unit $3$ - dimensional cube as $\Omega$ (Fig. 3):
				
		\begin{equation}\label{cube_3D}
			\begin{gathered}
				\Pi_{3}=\{t=(t_1,t_2,t_{3}) \in\mathcal {L}_{3} : t_i=0,1, i=1,2,3\}=\\		
				\{ \{{\omega}^0, \omega^1,...,\omega^{7} \} , 
				{\omega}^0=\tau^0,  	{\omega}^1=\tau^1, 	{\omega}^2=\tau^{L_1},
				{\omega}^3=\tau^{L_1+1}, 	{\omega}^4=\tau^{L_1 L_2}, \\
				{\omega}^5=\tau^{L_1 L_2 +1}, 	{\omega}^6=\tau^{L_1 L_2+L_1},	
				{\omega}^7=\tau^{L_1 L_2+L_1+1}=	{\omega}^{max}={\omega}^{r} \}.
			\end{gathered} 
		\end{equation}
		
		Then $ \Phi =\{{\omega}^0, \omega^2, \omega^4, \omega^6 \}$.		
	
		The Hamiltonian of the model has the form (\ref{Hamiltonian1}) ,		 
		where $\tau^i=(\tau_1^i,\tau_2^i,\tau_{3}^i) \in  \mathcal {L}_{3}  .  $

		\begin{picture}(180,200) 		
			\put(20,55){\line(1,0){100}}
			\put(20,55){\line(0,1){100}}
			\put(20,55){\circle*{3}}
			\put(120,55){\circle*{3}}
			\put(0,40){${\tau^0}={\omega^0}$}
			\put(0,160){${\tau^{L_1 L_2}}={\omega^4}$}
			\put(125,147){${\tau^{L_1 L_2+1}}={\omega^5}$}		 		\put(100,40){${\tau^1}={{\omega^1}}$}		
			\put(20,155){\line(1,0){100}}
			\put(20,55){\line(3,1){40}}
			\put(20,155){\line(3,1){40}}
			\put(120,155){\line(3,1){40}}
			\put(60,70){\line(0,1){100}}
			\put(60,69){\line(1,0){100}}
			\put(60,169){\line(1,0){100}}
			\put(160,69){\line(0,1){100}}	 	
			\put(120,55){\line(3,1){40}}
			\put(120,55){\line(0,1){100}}
			\put(120,55){\line(1,0){100}}
			\put(60,69){\line(1,0){100}}
			\put(60,69){\circle*{3}}
			\put(60,169){\circle*{3}}
			\put(160,69){\circle*{3}}
			\put(160,169){\circle*{3}}		
			\put(50,73){${\tau^{L_1}}={\omega^2}$}
			\put(50,173){${\tau^{L_1 L_2+L_1}}={\omega^6}$}
			\put(150,73){${\tau^{L_1+1}}={\omega^3}$}
			\put(150,173){${\tau^{L_1 L_2+L_1+1}}={\omega^7}$}
			\put(20,155){\circle*{3}}
			\put(120,155){\circle*{3}}
			\put(215,40){${\tau^2}$}
			\put(220,55){\circle*{3}}
			\qbezier[460](220,55)(520,1)(60,69)		
			\put(55,10){Fig. 3}
		\end{picture}
			
		Consequently,
		
		\begin{equation}\label{3D_eigenvector_1}
			\begin{gathered} 
				w_{ (\sigma_{\omega^0},\sigma_{\omega^{1}},\sigma_{\omega^{2}},
					\sigma_{\omega^{3}},\sigma_{\omega^{4}},\sigma_{\omega^{5}},
					\sigma_{\omega^{6}})  }(\sigma_{\omega^{0}}=\sigma_{\phi^{1}}, \sigma_{\omega^{2}}=\sigma_{\phi^{2}},
				\sigma_{\omega^{4}}=\sigma_{\phi^{3}},\sigma_{\omega^{6}}=\sigma_{\phi^{4}}) =\\
				(((W_0  (1 + \sigma_{\omega^0}) / 2 + 
				W_1 (1 - \sigma_{\omega^0}) / 2) (1 + \sigma_{\omega^2}) / 2 +\\
				(W_2 (1 + \sigma_{\omega^0}) / 2 +
				W_3 (1 - \sigma_{\omega^0}) / 2) (1 - \sigma_{\omega^2}) / 2) (1 + \sigma_{\omega^4}) / 2 + \\
				((W_4 (1 + \sigma_{\omega^0}) / 2 + W_5 (1 - \sigma_{\omega^0}) / 2) (1 + \sigma_{\omega^2}) / 2 + \\
				(W_6 (1 + \sigma_{\omega^0}) / 2 + 
				W_7 (1 - \sigma_{\omega^0}) / 2) (1 - \sigma_{\omega^2}) / 2) (1 - \sigma_{\omega^4}) / 2) (1 + \sigma_{\omega^6}) / 2 +\\	 		
				(((W_8 (1 + \sigma_{\omega^0}) / 2 + W_9 (1 - \sigma_{\omega^0}) / 2) (1 + \sigma_{\omega^2}) / 2 + \\
				(W_{10} (1 + \sigma_{\omega^0}) / 2 + 
				W_{11} (1 - \sigma_{\omega^0}) / 2) (1 - \sigma_{\omega^2}) / 2) (1 + \sigma_{\omega^4}) / 2 + \\
				((W_{12} (1 + \sigma_{\omega^0}) / 2 + 
				W_{13} (1 - \sigma_{\omega^0}) / 2) (1 + \sigma_{\omega^2}) / 2 +\\
				(W_{14} (1 + \sigma_{\omega^0}) / 2 +
				W_{15} (1 - \sigma_{\omega^0}) / 2) (1 - \sigma_{\omega^2}) / 2) (1 - \sigma_{\omega^4}) / 2) (1 - \sigma_{\omega^6}) / 2
				. \\					    			
			\end{gathered} 	
		\end{equation}	
			
		It is convenient to take $ W_0 = 1. $ as the normalization of the eigenvector $ \overrightarrow {V} $.
		
		The system of equations (\ref{main_system_small_2}) for this case can be written in the form 	
		 
		\begin{equation}\label{main_system_small_2_3D}%{Eigenvalue_max_omega2_3D}
			\begin{gathered} 
				{\lambda_{max}}		w_{ (\sigma_{\omega^0},\sigma_{\omega^{1}},\sigma_{\omega^{2}},
					\sigma_{\omega^{3}},\sigma_{\omega^{4}},\sigma_{\omega^{5}},
					\sigma_{\omega^{6}})  } =\\
				\sum_{\sigma_{\omega^{7}}} 		
				\exp(  \sum_{\{\omega^{j_1},\omega^{j_2},...,\omega^{j_s}\}\subset \Omega_{\tau^0}}
				K_{\omega^{j_1} , \omega^{j_2} , ..., \omega^{j_s}}\sigma_{\omega^{j_1}}\sigma_{\omega^{j_2}}...\sigma_{\omega^{j_s}})\\
				w_{ (\sigma_{\omega^1},\sigma_{\omega^{2}},\sigma_{\omega^{3}},
					\sigma_{\omega^{4}},\sigma_{\omega^{5}},\sigma_{\omega^{6}},
					\sigma_{\omega^{7}})  }			.		    			
			\end{gathered} 	
		\end{equation}
		
		We write the solution of the system (\ref{K_nuD}) for the 3D case in the following form

		\begin{equation}\label{K_3D}
			\begin{gathered}			
				K_{\omega^{j_1} , \omega^{j_2} , ..., \omega^{j_s}}=\\
				-(\sum_{\{\sigma_{\omega^0},\sigma_{\omega^1},...,\sigma_{\omega^{6}}\}}
				\sigma_{\omega^{j_1}}\sigma_{\omega^{j_2}}...\sigma_{\omega^{j_s}}	
				A_{\sigma_{\omega^0} ,\sigma_{\omega^1},...,\sigma_{\omega^{6}} })/2^{7},\\
				\{\sigma_{\omega^0} ,\sigma_{\omega^1},...,\sigma_{\omega^{6}} \} 	\neq
				(+1,+1,...,+1 )	, 			
			\end{gathered} 	
		\end{equation}	
		
		where 
	
		\begin{equation}\label{A_3D}
			\begin{gathered}
				A_{\sigma_{\omega^0} ,\sigma_{\omega^1},...,\sigma_{\omega^{6}} }=\\
				\ln(			\sum_{\sigma_{\omega^{7}}} 		
				\exp(  \sum_{\{\omega^{j_1},\omega^{j_2},...,\omega^{7}\}\subset \Omega_{\tau^0}}
				K_{\omega^{j_1} , \omega^{j_2} , ..., \omega^{7}}
				\sigma_{\omega^{j_1}}\sigma_{\omega^{j_2}}...
				\sigma_{\omega^{7}})	\cdot \\	
				w_{ (\sigma_{\omega^{1}},\sigma_{\omega^{2}},
					\sigma_{\omega^{3}},\sigma_{\omega^{4}},\sigma_{\omega^{5}},
					\sigma_{\omega^{6}},\sigma_{\omega^7})  }
				)
				- \\
				\ln(			\sum_{\sigma_{\omega^{7}}} 		
				\exp(  \sum_{\{\omega^{j_1},\omega^{j_2},...,\omega^{7}\}\subset \Omega_{\tau^0}}
				K_{\omega^{j_1} , \omega^{j_2} , ..., \omega^{7}}\sigma_{\omega^{7}}) \cdot\\
				w_{ (\sigma_{\omega^1}=+1,\sigma_{\omega^{2}}=+1,...,,\sigma_{\omega^{6}}=+1,\sigma_{\omega^{7}})})  +\\
				\ln(	w_{ (\sigma_{\omega^0}=+1,\sigma_{\omega^{1}}=+1,...,\sigma_{\omega^{6}}=+1)  })- 	
				\ln(w_{ (\sigma_{\omega^0},\sigma_{\omega^{1}},\sigma_{\omega^{2}},
					\sigma_{\omega^{3}},\sigma_{\omega^{4}},\sigma_{\omega^{5}},
					\sigma_{\omega^{6}})  }  ).
			\end{gathered} 	
		\end{equation}
	
				\textbf{	Example{ Numerical calculation of multiplicative coefficients and free energy for cube  model on the 3D lattice, general case}}\label{3D lattice}
			
		We define the free $128 $ parameters $	K_{\omega^{i_1} ,...,\omega^{i_s},\omega^{7}} $,
		, containing  $\omega^{7} $
	  , and  $W_i=i+1, i=0,...,15 $. Total unordered solution depends on $ 144 $ parameters
		Other $128$ parameters $	K_{\omega^{i_1} ,...,\omega^{i_s}} $, not containing 
		$\omega^{7}, $
		are calculated by the formulas  (\ref{K_3D})-(\ref{A_3D}) and are in the last column of the table. 		
		
		\begin{table}
			\caption{Multi-spin interaction coefficients for cube model on the 3D lattice, general case}
			\label{table_1}
		\begin{tabular}{ | l | l | l | l | }			
			\hline
			${ No.}$ & $ \{{i_1} ,...,{i_s}\}$ & $	K_{\omega^{i_1} ,...,\omega^{i_s},\omega^{7}} $& $	K_{\omega^{i_1} ,...,\omega^{i_s}}$ \\ \hline
			0  & $ \varnothing $  &    -0.9590    &    0.000000000000   \\    
			1  &  6   &    -0.5330    &    -0.379046232278   \\    
			2  &  5   &    -0.6660    &    0.142211623068   \\    
			3  &  5 6   &    -0.5000    &    -0.379518097153   \\    
			4  &  4   &    0.1690    &    0.418665068038   \\    
			5  &  4 6   &    0.7240    &    -0.015571879954   \\    
			6  &  4 5   &    0.4780    &    -0.191009769175   \\    
			7  &  4 5 6   &    0.3580    &    0.552160328153   \\    
			8  &  3   &    -0.0380    &    0.276243393364   \\    
			9  &  3 6   &    -0.5360    &    0.382664591954   \\    
			10  &  3 5   &    0.7050    &    0.284722589540   \\    
			11  &  3 5 6   &    -0.8550    &    -0.953293690482   \\    
			12  &  3 4   &    0.2810    &    0.127881283934   \\    
			13  &  3 4 6   &    -0.1730    &    -0.546815648493   \\    
			14  &  3 4 5   &    0.9610    &    -0.160889227814   \\    
			15  &  3 4 5 6   &    -0.5090    &    0.208902503053   \\    
			16  &  2   &    -0.0050    &    -0.403697533293   \\    
			17  &  2 6   &    0.9420    &    0.215498806729   \\    
			18  &  2 5   &    -0.1730    &    0.448447160140   \\    
			19  &  2 5 6   &    0.4360    &    0.066243725686   \\    
			20  &  2 4   &    -0.6090    &    -0.091438717858   \\    
			21  &  2 4 6   &    -0.3960    &    0.000142820253   \\    
			22  &  2 4 5   &    0.9020    &    0.037012053881   \\    
			23  &  2 4 5 6   &    -0.8470    &    0.033179886870   \\    
			24  &  2 3   &    -0.7080    &    -0.309608957576   \\    
			25  &  2 3 6   &    -0.6180    &    0.190834307340   \\    
			26  &  2 3 5   &    0.4210    &    0.027291467420   \\    
			27  &  2 3 5 6   &    -0.2840    &    -0.422177722572   \\    
			28  &  2 3 4   &    0.7180    &    0.070737925480   \\    
			29  &  2 3 4 6   &    0.8950    &    0.345317928814   \\    
			30  &  2 3 4 5   &    0.4470    &    0.896734626961   \\    
			31  &  2 3 4 5 6   &    0.7260    &    0.450967179463   \\    
			32  &  1   &    -0.2290    &    -0.252947812800   \\    
			33  &  1 6   &    0.5380    &    -0.111424106275   \\    
			34  &  1 5   &    0.8690    &    0.245696675844   \\    
			35  &  1 5 6   &    0.9120    &    0.446338504569   \\    
			36  &  1 4   &    0.6670    &    0.215167461642   \\    
			37  &  1 4 6   &    -0.7010    &    -0.357213820067   \\    
			38  &  1 4 5   &    0.0350    &    0.025998977020   \\    
			39  &  1 4 5 6   &    0.8940    &    0.454040219232   \\    
			40  &  1 3   &    -0.2970    &    0.045720414878   \\    
			41  &  1 3 6   &    0.8110    &    -0.016378972164   \\    
			42  &  1 3 5   &    0.3220    &    0.576276729173   \\    
			\hline
		\end{tabular}
			\end{table}

		\begin{tabular}{ | l | l | l | l | }
			\hline
			${ No.}$ &  $ \{{i_1} ,...,{i_s}\}$ & $	K_{\omega^{i_1} ,...,\omega^{i_s},\omega^{7}} $& $	K_{\omega^{i_1} ,...,\omega^{i_s}}$  \\ \hline
			43  &  1 3 5 6   &    -0.6670    &    -0.085103136065   \\    
			44  &  1 3 4   &    0.6730    &    0.260333650275   \\    
			45  &  1 3 4 6   &    -0.3360    &    0.018402858556   \\    
			46  &  1 3 4 5   &    0.1410    &    -0.228737248646   \\    
			47  &  1 3 4 5 6   &    0.7110    &    0.181295223479   \\    
			48  &  1 2   &    -0.7470    &    0.192227239409   \\    
			49  &  1 2 6   &    -0.1320    &    -0.184582953850   \\    
			50  &  1 2 5   &    0.5470    &    -0.133470739538   \\    
			51  &  1 2 5 6   &    0.6440    &    0.143449561581   \\    
			52  &  1 2 4   &    -0.3380    &    -0.001992414255   \\    
			53  &  1 2 4 6   &    -0.2430    &    -0.086762892607   \\    
			54  &  1 2 4 5   &    -0.9630    &    0.142378206874   \\    
			55  &  1 2 4 5 6   &    -0.1410    &    -0.631511160521   \\    
			56  &  1 2 3   &    -0.2770    &    0.419819514087   \\    
			57  &  1 2 3 6   &    0.7410    &    -0.257286850158   \\    
			58  &  1 2 3 5   &    0.5290    &    0.317340798941   \\    
			59  &  1 2 3 5 6   &    -0.2220    &    0.539226683743   \\    
			60  &  1 2 3 4   &    -0.6840    &    -0.572254314148   \\    
			61  &  1 2 3 4 6   &    0.0350    &    -0.508124003229   \\    
			62  &  1 2 3 4 5   &    -0.8100    &    0.083471878534   \\    
			63  &  1 2 3 4 5 6   &    0.8420    &    -0.512569609229   \\    
			64  &  0   &    -0.7120    &    -0.304070758119   \\    
			65  &  0 6   &    -0.8940    &    -0.174846482917   \\    
			66  &  0 5   &    0.0400    &    0.032907610556   \\    
			67  &  0 5 6   &    -0.0580    &    -0.179042204018   \\    
			68  &  0 4   &    0.2640    &    0.651767718020   \\    
			69  &  0 4 6   &    -0.3520    &    -0.382313874261   \\    
			70  &  0 4 5   &    0.4460    &    -0.019201981635   \\    
			71  &  0 4 5 6   &    0.8050    &    0.743440233816   \\    
			72  &  0 3   &    0.8900    &    0.325563068418   \\    
			73  &  0 3 6   &    -0.2710    &    -0.269863102161   \\    
			74  &  0 3 5   &    -0.6300    &    -0.459104744937   \\    
			75  &  0 3 5 6   &    0.3500    &    -0.133757061751   \\    
			76  &  0 3 4   &    0.0060    &    -0.224023321885   \\    
			77  &  0 3 4 6   &    0.1010    &    0.130384139367   \\    
			78  &  0 3 4 5   &    -0.6070    &    -0.362346585020   \\    
			79  &  0 3 4 5 6   &    0.5480    &    0.925937476851   \\    
			80  &  0 2   &    0.6290    &    -0.160817421538   \\    
			81  &  0 2 6   &    -0.3770    &    -0.129635535155   \\    
			82  &  0 2 5   &    -0.9160    &    -0.021187519794   \\    
			83  &  0 2 5 6   &    0.9540    &    0.303671222289   \\    
			84  &  0 2 4   &    -0.2440    &    -0.236235921559   \\    
			85  &  0 2 4 6   &    0.8400    &    0.021138654751   \\    
			86  &  0 2 4 5   &    -0.0340    &    -0.367373554751   \\    
			\hline
		\end{tabular}

		\begin{tabular}{ | l | l | l | l | }
			\hline
			${ No.}$ &  $ \{{i_1} ,...,{i_s}\}$ &  $	K_{\omega^{i_1} ,...,\omega^{i_s},\omega^{7}} $ &  $	K_{\omega^{i_1} ,...,\omega^{i_s}}$  \\ \hline
			87  &  0 2 4 5 6   &    0.3760    &    -0.160269836991   \\    
			88  &  0 2 3   &    0.9310    &    0.386961845874   \\    
			89  &  0 2 3 6   &    -0.6920    &    -0.999506904306   \\    
			90  &  0 2 3 5   &    -0.0560    &    -0.266726413649   \\    
			91  &  0 2 3 5 6   &    -0.5610    &    -0.389832355853   \\    
			92  &  0 2 3 4   &    -0.3740    &    0.589685176802   \\    
			93  &  0 2 3 4 6   &    0.3230    &    -0.136503598510   \\    
			94  &  0 2 3 4 5   &    0.5370    &    0.252986328530   \\    
			95  &  0 2 3 4 5 6   &    0.5380    &    0.561101185654   \\    
			96  &  0 1   &    -0.8820    &    -0.614494938614   \\    
			97  &  0 1 6   &    -0.9180    &    0.094186215087   \\    
			98  &  0 1 5   &    -0.0710    &    -0.495480319115   \\    
			99  &  0 1 5 6   &    -0.4590    &    0.278818413718   \\    
			100  &  0 1 4   &    -0.1670    &    -0.353921515032   \\    
			101  &  0 1 4 6   &    0.1150    &    -0.208859237885   \\    
			102  &  0 1 4 5   &    -0.3610    &    0.415497288673   \\    
			103  &  0 1 4 5 6   &    0.6580    &    0.184005076081   \\    
			104  &  0 1 3   &    -0.2960    &    0.231884125165   \\    
			105  &  0 1 3 6   &    0.9300    &    -0.297705503151   \\    
			106  &  0 1 3 5   &    0.9770    &    -0.122743994152   \\    
			107  &  0 1 3 5 6   &    -0.6940    &    -0.040893126584   \\    
			108  &  0 1 3 4   &    0.6730    &    -0.137055273136   \\    
			109  &  0 1 3 4 6   &    -0.6140    &    0.517721807397   \\    
			110  &  0 1 3 4 5   &    0.0210    &    0.247575542266   \\    
			111  &  0 1 3 4 5 6   &    -0.2550    &    -0.408994250272   \\    
			112  &  0 1 2   &    -0.0760    &    0.000143283135   \\    
			113  &  0 1 2 6   &    0.0720    &    -0.189772729402   \\    
			114  &  0 1 2 5   &    -0.7300    &    -0.057641009793   \\    
			115  &  0 1 2 5 6   &    0.8290    &    0.377304965015   \\    
			116  &  0 1 2 4   &    -0.2230    &    -0.896982823455   \\    
			117  &  0 1 2 4 6   &    0.5730    &    0.005295252682   \\    
			118  &  0 1 2 4 5   &    0.0970    &    -0.578653391698   \\    
			119  &  0 1 2 4 5 6   &    -0.4880    &    -0.173186217601   \\    
			120  &  0 1 2 3   &    0.9860    &    0.407624202508   \\    
			121  &  0 1 2 3 6   &    0.2900    &    -0.333515770198   \\    
			122  &  0 1 2 3 5   &    0.1610    &    -0.307969720110   \\    
			123  &  0 1 2 3 5 6   &    -0.3640    &    -0.010260842557   \\    
			124  &  0 1 2 3 4   &    -0.6450    &    -0.047631120149   \\    
			125  &  0 1 2 3 4 6   &    -0.2330    &    0.149222316210   \\    
			126  &  0 1 2 3 4 5   &    0.6550    &    0.726174894060   \\    
			127  &  0 1 2 3 4 5 6   &    0.5740    &    -0.105426815869   \\    
			\hline
		\end{tabular}
	
		Wherein 
		\begin{equation}\label{3D_calculate}
			\begin{gathered}	
				\lambda_{max}= 201.511609971965 , \;\;
				- f/ {(kT)}= 5.305846997454 ,
			\end{gathered} 	
		\end{equation}
		where $f$ is free energy per one lattice site.
	
		\subsection{Free energy  for cube  models invariant under the reversal of signs of all spins on the 3D lattice}\label{3D_cube_even}
		\hfill \break
				
		Let us consider the unit $3$ - dimensional cube as $\Omega$ (Fig. 3) (\ref{cube_3D}):		
		
		Then $ \Phi =\{{\omega}^0, \omega^2, \omega^4, \omega^6 \}$.
		
		The Hamiltonian of the model has the form (\ref{Hamiltonian1}) , where $s=2 p$, that is, only the coefficients are nonzero $	K_{\omega^{j_1} , \omega^{j_2} , ..., \omega^{j_s}}$ with even $s$.
		
		The elementary transfer matrix
		and the eigenvector are centrally symmetrical, so
		 		\begin{equation}\label{3D_eigenvector_even}
			\begin{gathered} 
				w_{ (\sigma_{\omega^0},\sigma_{\omega^{1}},\sigma_{\omega^{2}},
					\sigma_{\omega^{3}},\sigma_{\omega^{4}},\sigma_{\omega^{5}},
					\sigma_{\omega^{6}})  }(\sigma_{\omega^{0}}=\sigma_{\phi^{1}}, \sigma_{\omega^{2}}=\sigma_{\phi^{2}},
				\sigma_{\omega^{4}}=\sigma_{\phi^{3}},\sigma_{\omega^{6}}=\sigma_{\phi^{4}}) =\\
				(((W_0  (1 + \sigma_{\omega^0}) / 2 + 
				W_1 (1 - \sigma_{\omega^0}) / 2) (1 + \sigma_{\omega^2}) / 2 +\\
				(W_2 (1 + \sigma_{\omega^0}) / 2 +
				W_3 (1 - \sigma_{\omega^0}) / 2) (1 - \sigma_{\omega^2}) / 2) (1 + \sigma_{\omega^4}) / 2 + \\
				((W_4 (1 + \sigma_{\omega^0}) / 2 + W_5 (1 - \sigma_{\omega^0}) / 2) (1 + \sigma_{\omega^2}) / 2 + \\
				(W_6 (1 + \sigma_{\omega^0}) / 2 + 
				W_7 (1 - \sigma_{\omega^0}) / 2) (1 - \sigma_{\omega^2}) / 2) (1 - \sigma_{\omega^4}) / 2) (1 + \sigma_{\omega^6}) / 2 +\\	 		
				(((W_7 (1 + \sigma_{\omega^0}) / 2 + W_6 (1 - \sigma_{\omega^0}) / 2) (1 + \sigma_{\omega^2}) / 2 + \\
				(W_{5} (1 + \sigma_{\omega^0}) / 2 + 
				W_{4} (1 - \sigma_{\omega^0}) / 2) (1 - \sigma_{\omega^2}) / 2) (1 + \sigma_{\omega^4}) / 2 + \\
				((W_{3} (1 + \sigma_{\omega^0}) / 2 + 
				W_{2} (1 - \sigma_{\omega^0}) / 2) (1 + \sigma_{\omega^2}) / 2 +\\
				(W_{1} (1 + \sigma_{\omega^0}) / 2 +
				W_0 (1 - \sigma_{\omega^0}) / 2) (1 - \sigma_{\omega^2}) / 2) (1 - \sigma_{\omega^4}) / 2) (1 - \sigma_{\omega^6}) / 2
				. \\					    			
			\end{gathered} 	
		\end{equation}

	We can simplify the formula (\ref{K_even}) for this case:
	
	%%%%%%%%%%%%%%%%%%%%%%%%%%
		\begin{equation}\label{K_even3D_cube}
		\begin{gathered}			
			K_{\omega^{j_1} , \omega^{j_2} , ..., \omega^{j_{2 p}}}=\\
			-(\sum_{\{\sigma_{\omega^0},\sigma_{\omega^1},...,\sigma_{\omega^{5}}\}}
			\sigma_{\omega^{j_1}}\sigma_{\omega^{j_2}}...\sigma_{\omega^{j_{2 p}}}	
			A_{\sigma_{\omega^0} ,\sigma_{\omega^1},...,\sigma_{\omega^{5}},
				\sigma_{\omega^{6}}=+1 })/2^{6},\\  			
			\{\sigma_{\omega^0} ,\sigma_{\omega^1},...,\sigma_{\omega^{6}} \} 	\neq
			(+1,+1,...,+1 )	,
			{\{\omega^{j_1},\omega^{j_2},...,\omega^{j_{2 p}}\}\subset \Omega'_{\tau^0}},\\ 	
			K_{\omega^{j_1} , \omega^{j_2} , ..., \omega^{j_{2 p-1}}}=	0,\\
			{\{\omega^{j_1},\omega^{j_2},...,\omega^{j_{2 p -1}}\}\subset \Omega'_{\tau^0}},	  				
		\end{gathered} 	
	\end{equation}
	
	where   $	A_{\sigma_{\omega^0} ,\sigma_{\omega^1},...,\sigma_{\omega^{5}},
		\sigma_{\omega^{6}}=+1 } $ we find by the formulas (\ref{A_right}) with a centrally symmetric vector $	\overrightarrow{ w}$ (\ref{3D_eigenvector_even}).

	%%%%%%%%%%%%%%%%%%%%%%%%	
	\textbf{	Example{ 1. Numerical calculation of multiplicative coefficients and free energy for cube  model invariant under the reversal of signs of all spins on the 3D lattice}.}\label{example_3D_even_1}

		We define the free $64 $ parameters $	K_{\omega^{i_1} ,...,\omega^{i_{2 p - 1}},\omega^{7}} $,
		, containing $\omega^{7} $
		(see table \ref{table_2} ), and  $W_i=i+1, W_{15-i}=W_{i}, i=0,...,7, $  . The total disordered solution depends on $71=64+7$ parameters.
		Other $128$ parameters $	K_{\omega^{i_1} ,...,\omega^{i_s}} $, not containing
		$\omega^{7}, $
		are calculated by the formulas (\ref{K_3D})-(\ref{A_3D}) and are in the last column of the table. Moreover, it is clear that the coefficients $K_{\omega^{i_1} ,...,\omega^{i_s}}$, corresponding to the interaction of products of an odd number of spins are equal to zero.

		\begin{table}
			\caption{	Multi-spin interaction coefficients for cube model invariant under the reversal of signs of all spins on the 3D lattice	
		}
			\label{table_2}
		\begin{tabular}{ | l | l | l | l | }
			\hline
			${ No.}$ & $ \{{i_1} ,...,{i_s}\}$ & $	K_{\omega^{i_1} ,...,\omega^{i_s},\omega^{7}} $& $	K_{\omega^{i_1} ,...,\omega^{i_s}}$ \\ \hline    
			0  & $ \varnothing $  &    0.0000    &    0.000000000000   \\    
			1  &  6   &    -0.5330    &    0.000000000000   \\    
			2  &  5   &    -0.6660    &    0.000000000000   \\    
			3  &  5 6   &    0.0000    &    -0.011955361081   \\    
			4  &  4   &    0.1690    &    0.000000000000   \\    
			5  &  4 6   &    0.0000    &    -0.562146987484   \\    
			6  &  4 5   &    0.0000    &    -0.096967600783   \\    
			7  &  4 5 6   &    0.3580    &    0.000000000000   \\    
			8  &  3   &    -0.0380    &    0.000000000000   \\    
			9  &  3 6   &    0.0000    &    -0.427953554050   \\    
			10  &  3 5   &    0.0000    &    -0.076938164059   \\    
			11  &  3 5 6   &    -0.8550    &    0.000000000000   \\    
			12  &  3 4   &    0.0000    &    0.324394377158   \\    
			13  &  3 4 6   &    -0.1730    &    0.000000000000   \\    
			14  &  3 4 5   &    0.9610    &    0.000000000000   \\    
			15  &  3 4 5 6   &    0.0000    &    0.313654916461   \\    
			16  &  2   &    -0.0050    &    0.000000000000   \\    
			17  &  2 6   &    0.0000    &    -0.480347581396   \\    
			18  &  2 5   &    0.0000    &    0.274036043739   \\    
			19  &  2 5 6   &    0.4360    &    0.000000000000   \\    
			20  &  2 4   &    0.0000    &    0.406368117011   \\    
			21  &  2 4 6   &    -0.3960    &    0.000000000000   \\    
			22  &  2 4 5   &    0.9020    &    0.000000000000   \\    
			23  &  2 4 5 6   &    0.0000    &    0.604766952407   \\    
			24  &  2 3   &    0.0000    &    0.165441839964   \\    
			25  &  2 3 6   &    -0.6180    &    0.000000000000   \\    
			26  &  2 3 5   &    0.4210    &    0.000000000000   \\    
			27  &  2 3 5 6   &    0.0000    &    -0.027989568441   \\    
			28  &  2 3 4   &    0.7180    &    0.000000000000   \\    
			29  &  2 3 4 6   &    0.0000    &    0.631861607451   \\    
			30  &  2 3 4 5   &    0.0000    &    0.402919149714   \\    
			31  &  2 3 4 5 6   &    0.7260    &    0.000000000000   \\    
			32  &  1   &    -0.2290    &    0.000000000000   \\    
			33  &  1 6   &    0.0000    &    -0.062998113769   \\    
			34  &  1 5   &    0.0000    &    0.420876617024   \\    
			35  &  1 5 6   &    0.9120    &    0.000000000000   \\    
			36  &  1 4   &    0.0000    &    0.291322247779   \\    
			37  &  1 4 6   &    -0.7010    &    0.000000000000   \\    
			38  &  1 4 5   &    0.0350    &    0.000000000000   \\    
			39  &  1 4 5 6   &    0.0000    &    -0.184583633798   \\    
			40  &  1 3   &    0.0000    &    0.147645302717   \\    
			41  &  1 3 6   &    0.8110    &    0.000000000000   \\    
			42  &  1 3 5   &    0.3220    &    0.000000000000   \\     
			\hline
		\end{tabular}
			\end{table}

		\begin{tabular}{ | l | l | l | l | }
			\hline
			${ No.}$ &  $ \{{i_1} ,...,{i_s}\}$ & $	K_{\omega^{i_1} ,...,\omega^{i_s},\omega^{7}} $& $	K_{\omega^{i_1} ,...,\omega^{i_s}}$  \\ \hline
			43  &  1 3 5 6   &    0.0000    &    0.070296374959   \\    
			44  &  1 3 4   &    0.6730    &    0.000000000000   \\    
			45  &  1 3 4 6   &    0.0000    &    -0.207232986725   \\    
			46  &  1 3 4 5   &    0.0000    &    0.082781999276   \\    
			47  &  1 3 4 5 6   &    0.7110    &    0.000000000000   \\    
			48  &  1 2   &    0.0000    &    -0.102661035637   \\    
			49  &  1 2 6   &    -0.1320    &    0.000000000000   \\    
			50  &  1 2 5   &    0.5470    &    0.000000000000   \\    
			51  &  1 2 5 6   &    0.0000    &    0.074309123798   \\    
			52  &  1 2 4   &    -0.3380    &    0.000000000000   \\    
			53  &  1 2 4 6   &    0.0000    &    -0.067189718198   \\    
			54  &  1 2 4 5   &    0.0000    &    0.092130587872   \\    
			55  &  1 2 4 5 6   &    -0.1410    &    0.000000000000   \\    
			56  &  1 2 3   &    -0.2770    &    0.000000000000   \\    
			57  &  1 2 3 6   &    0.0000    &    -0.058819998844   \\    
			58  &  1 2 3 5   &    0.0000    &    0.071580968072   \\    
			59  &  1 2 3 5 6   &    -0.2220    &    0.000000000000   \\    
			60  &  1 2 3 4   &    0.0000    &    -0.501947937498   \\    
			61  &  1 2 3 4 6   &    0.0350    &    0.000000000000   \\    
			62  &  1 2 3 4 5   &    -0.8100    &    0.000000000000   \\    
			63  &  1 2 3 4 5 6   &    0.0000    &    -0.330886336628   \\    
			64  &  0   &    -0.7120    &    0.000000000000   \\    
			65  &  0 6   &    0.0000    &    -0.261549695481   \\    
			66  &  0 5   &    0.0000    &    0.151963865704   \\    
			67  &  0 5 6   &    -0.0580    &    0.000000000000   \\    
			68  &  0 4   &    0.0000    &    0.134805759710   \\    
			69  &  0 4 6   &    -0.3520    &    0.000000000000   \\    
			70  &  0 4 5   &    0.4460    &    0.000000000000   \\    
			71  &  0 4 5 6   &    0.0000    &    0.140179562678   \\    
			72  &  0 3   &    0.0000    &    -0.193267159756   \\    
			73  &  0 3 6   &    -0.2710    &    0.000000000000   \\    
			74  &  0 3 5   &    -0.6300    &    0.000000000000   \\    
			75  &  0 3 5 6   &    0.0000    &    -0.368169038652   \\    
			76  &  0 3 4   &    0.0060    &    0.000000000000   \\    
			77  &  0 3 4 6   &    0.0000    &    0.647698107704   \\    
			78  &  0 3 4 5   &    0.0000    &    0.366504814660   \\    
			79  &  0 3 4 5 6   &    0.5480    &    0.000000000000   \\    
			80  &  0 2   &    0.0000    &    -0.757754596316   \\    
			81  &  0 2 6   &    -0.3770    &    0.000000000000   \\    
			82  &  0 2 5   &    -0.9160    &    0.000000000000   \\    
			83  &  0 2 5 6   &    0.0000    &    0.165682725282   \\    
			84  &  0 2 4   &    -0.2440    &    0.000000000000   \\    
			85  &  0 2 4 6   &    0.0000    &    0.122308936257   \\    
			86  &  0 2 4 5   &    0.0000    &    0.085098077100   \\    
			\hline
		\end{tabular}

		\begin{tabular}{ | l | l | l | l | }
			\hline
			${ No.}$ &  $ \{{i_1} ,...,{i_s}\}$ &  $	K_{\omega^{i_1} ,...,\omega^{i_s},\omega^{7}} $ &  $	K_{\omega^{i_1} ,...,\omega^{i_s}}$  \\ \hline
			87  &  0 2 4 5 6   &    0.3760    &    0.000000000000   \\    
			88  &  0 2 3   &    0.9310    &    0.000000000000   \\    
			89  &  0 2 3 6   &    0.0000    &    -0.544237144556   \\    
			90  &  0 2 3 5   &    0.0000    &    0.076735321209   \\    
			91  &  0 2 3 5 6   &    -0.5610    &    0.000000000000   \\    
			92  &  0 2 3 4   &    0.0000    &    0.085304214236   \\    
			93  &  0 2 3 4 6   &    0.3230    &    0.000000000000   \\    
			94  &  0 2 3 4 5   &    0.5370    &    0.000000000000   \\    
			95  &  0 2 3 4 5 6   &    0.0000    &    0.197245979972   \\    
			96  &  0 1   &    0.0000    &    -0.200130357830   \\    
			97  &  0 1 6   &    -0.9180    &    0.000000000000   \\    
			98  &  0 1 5   &    -0.0710    &    0.000000000000   \\    
			99  &  0 1 5 6   &    0.0000    &    -0.042788834543   \\    
			100  &  0 1 4   &    -0.1670    &    0.000000000000   \\    
			101  &  0 1 4 6   &    0.0000    &    -0.232788489289   \\    
			102  &  0 1 4 5   &    0.0000    &    0.668321942891   \\    
			103  &  0 1 4 5 6   &    0.6580    &    0.000000000000   \\    
			104  &  0 1 3   &    -0.2960    &    0.000000000000   \\    
			105  &  0 1 3 6   &    0.0000    &    -0.535548941107   \\    
			106  &  0 1 3 5   &    0.0000    &    -0.081087609649   \\    
			107  &  0 1 3 5 6   &    -0.6940    &    0.000000000000   \\    
			108  &  0 1 3 4   &    0.0000    &    -0.024674644802   \\    
			109  &  0 1 3 4 6   &    -0.6140    &    0.000000000000   \\    
			110  &  0 1 3 4 5   &    0.0210    &    0.000000000000   \\    
			111  &  0 1 3 4 5 6   &    0.0000    &    0.033959300411   \\    
			112  &  0 1 2   &    -0.0760    &    0.000000000000   \\    
			113  &  0 1 2 6   &    0.0000    &    -0.141630091121   \\    
			114  &  0 1 2 5   &    0.0000    &    0.470174443338   \\    
			115  &  0 1 2 5 6   &    0.8290    &    0.000000000000   \\    
			116  &  0 1 2 4   &    0.0000    &    -0.641411560156   \\    
			117  &  0 1 2 4 6   &    0.5730    &    0.000000000000   \\    
			118  &  0 1 2 4 5   &    0.0970    &    0.000000000000   \\    
			119  &  0 1 2 4 5 6   &    0.0000    &    0.214733511320   \\    
			120  &  0 1 2 3   &    0.0000    &    0.526732963138   \\    
			121  &  0 1 2 3 6   &    0.2900    &    0.000000000000   \\    
			122  &  0 1 2 3 5   &    0.1610    &    0.000000000000   \\    
			123  &  0 1 2 3 5 6   &    0.0000    &    0.091406492260   \\    
			124  &  0 1 2 3 4   &    -0.6450    &    0.000000000000   \\    
			125  &  0 1 2 3 4 6   &    0.0000    &    0.305316969256   \\    
			126  &  0 1 2 3 4 5   &    0.0000    &    0.220271750679   \\    
			127  &  0 1 2 3 4 5 6   &    0.5740    &    0.000000000000 \\ 
			\hline
		\end{tabular}
	\\
		
		Wherein
		\begin{equation}\label{3D_lambda_even}
			\begin{gathered}	
				\lambda_{max}= 38.0895468788 , \;\;
				- f/ {(kT)}= 3.639939884353 ,
			\end{gathered} 	
		\end{equation}
		where $f$ is free energy per one lattice site.

			\textbf{	Example{ 2. Numerical calculation of multispin coefficients and free energy for cube  model invariant under the reversal of signs of all spins on the 3D lattice}, for three nonzero multispin parametric coefficients.}\label{example_3D_even_2}
			
			We define the free parameters:
		$  K_{\omega^3, \omega^7}= 1, 
  K_{\omega^5, \omega^7}= 1, 
	  K_{\omega^6, \omega^7}= 1, 	
		W_i=1, i=0, 2, 3, ..., 12, 13, 15,
		W_1=W_{14}=2 . $
		The rest of the free parameters are set equal to zero $	K_{\omega^{i_1} ,...,\omega^{i_s},\omega^{7}}=0. $
		
			Other $128$ parameters $	K_{\omega^{i_1} ,...,\omega^{i_s}} $, not containing
		$\omega^{7}, $
		are calculated by the formulas  (\ref{K_3D})-(\ref{A_3D}).
		Wherein

			\begin{equation}\label{3D_calculate_example2}
			\begin{gathered}
					K_{\omega^3, \omega^5 }= 
						K_{\omega^3, \omega^6 }=
				K_{\omega^5, \omega^6 }= -0.468886918524,\;\; 
				\lambda_{max}= 4.932346950116 , \\
				- f/ {(kT)}= 1.595814929567 ,
			\end{gathered} 	
		\end{equation}
		where $f$ is free energy per one lattice site.
		  Other coefficients $K_{\omega^{i_1} ,...,\omega^{i_s}} $, 
		  which do not contain $ \omega^{7}, $ are zeroed during the calculation.
	
	%%%%%%%%%%%%%%%%%%%%%%%%%%%%%%%%%%%%%%%%%%%%%%%%%%%%%%%%%%%%%%%%%%%%%%%%%%
	\section{Free energy  for $\nu$ - dimensional generalized models ANNNI}\label{nuD_ANNNI}	
	
	Let us consider the support of $\nu$ - dimensional model ANNNI as $\Omega$

	\begin{equation}\label{ANNNI_nu}
		\begin{gathered}
			\mathcal {A}_{\nu}=\{t=(t_1,t_2,...,t_{\nu}) \in\mathcal {L}_{\nu} : t_1=0,1,2 , \;\; t_i=0,1, i=2,3,...,\nu\}=\\		
			=\{{\omega}^0, \omega^1,...,\omega^{2^{\nu}+2^{\nu-1}-1} \} , \\
			{\omega}^0=\tau^0,  	{\omega}^1=\tau^1,   	{\omega}^2=\tau^2,
				{\omega}^3=\tau^{L_1}, 	{\omega}^4=\tau^{L_1+1},		{\omega}^5=\tau^{L_1+2},\\	
 		{\omega}^6=\tau^{L_1 L_2},
			{\omega}^7=\tau^{L_1 L_2 +1}, {\omega}^8=\tau^{L_1 L_2 +2},	{\omega}^9=\tau^{L_1 L_2+L_1},\\	
			{\omega}^{10}=\tau^{L_1 L_2+L_1+1}, 	{\omega}^{11}=\tau^{L_1 L_2+L_1+2},..., \\
			{\omega}^{2^{\nu}+2^{\nu-1}-1}=\tau^{L_1 L_2...L_{\nu -1}+L_1 L_2 ... L_{\nu -2}+...+L_1 L_2+L_1+2}=	{\omega}^{max}={\omega}^{r}.
		\end{gathered} 
	\end{equation}	
	
	Then	
	
	\begin{equation}\label{phi_ANNNI_nu}
		\begin{gathered}
			\Phi =\{{\omega}^0, \omega^1, \omega^3, \omega^4,{\omega}^6, \omega^7, \omega^9, \omega^{10},...,\omega^{2^{\nu}+2^{\nu-1}-2}\} .
		\end{gathered} 
	\end{equation}
	
	The Hamiltonian of the model has the form  (\ref{Hamiltonian1}) ,		 
	where $\tau^i=(\tau_1^i,\tau_2^i,...,\tau_{\nu}^i) \in  \mathcal {L}_{\nu}    $ 
	
	From (\ref{phi_ANNNI_nu}) the components of the eigenvector $	\overrightarrow{w}$  have the form

	\begin{equation}\label{w_ANNNI_nu}
		\begin{gathered} 
			w_{ (\sigma_{\omega^0},\sigma_{\omega^{1}},...,\sigma_{\omega^{{2^{\nu}}+
						{2^{\nu-1}}-2}})  }(\sigma_{\omega^{3 k - 3}}=\sigma_{\phi^{2 k - 1}},\sigma_{\omega^{3 k - 2}}=\sigma_{\phi^{2 k }}, k=1, 2, ..., {2^{\nu-1}}) =\\
			W_{ (\sigma_{\phi^1},\sigma_{\phi^{2}},...,\sigma_{\phi^{{2^{\nu}}}})}.\\		     			     			
		\end{gathered} 	
	\end{equation}

	It is convenient to normalize the eigenvector  $ 	\overrightarrow{w}$, taking  $	W_{0}=1.$

	The solution  (\ref{K}) for $\mathcal {A}_{\nu}$ can be written in the form

	\begin{equation}\label{K_ANNNI_nuD}
		\begin{gathered}			
			K_{\omega^{j_1} , \omega^{j_2} , ..., \omega^{j_s}}=\\
			-(\sum_{\{\sigma_{\omega^0},\sigma_{\omega^1},...,\sigma_{\omega^{2^{\nu}+2^{\nu-1}-2}}\}}
			\sigma_{\omega^{j_1}}\sigma_{\omega^{j_2}}...\sigma_{\omega^{j_s}}	
			A_{\sigma_{\omega^0} ,\sigma_{\omega^1},...,\sigma_{{\omega^{2^{\nu}+2^{\nu-1}-2}}} })/2^{({2^{\nu}+2^{\nu-1}-1})},\\
			\{\sigma_{\omega^0} ,\sigma_{\omega^1},...,\sigma_{\omega^{2^{\nu}+2^{\nu-1}-2}} \} 	\neq
			(+1,+1,...,+1 )	, 			
		\end{gathered} 	
	\end{equation}
		
	where 
	
	\begin{equation}\label{A_ANNNI_nuD}
		\begin{gathered}
			A_{\sigma_{\omega^0} ,\sigma_{\omega^1},...,\sigma_{\omega^{2^{\nu}+2^{\nu-1}-2}} }=\\
			\ln(			\sum_{\sigma_{\omega^{2^{\nu}+2^{\nu-1}-1}}} 		
			\exp(  \sum_{\{\omega^{j_1},\omega^{j_2},...,\omega^{2^{\nu}+2^{\nu-1}-1}\}\subset \Omega_{\tau^0}}
			K_{\omega^{j_1} , \omega^{j_2} , ..., \omega^{2^{\nu}+2^{\nu-1}-1}} \cdot\\
			\sigma_{\omega^{j_1}}\sigma_{\omega^{j_2}}...
			\sigma_{\omega^{2^{\nu}+2^{\nu-1}-1}}) \cdot \\		
			w_{ (\sigma_{\omega^{1}},\sigma_{\omega^{2}},...
				,
				\sigma_{\omega^{2^{\nu}+2^{\nu-1}-2}},\sigma_{\omega^{2^{\nu}+2^{\nu-1}-1}})  }
			)
			- \\
			\ln(			\sum_{\sigma_{\omega^{2^{\nu}+2^{\nu-1}-1}}} 		
			\exp(  \sum_{\{\omega^{j_1},\omega^{j_2},...,\omega^{2^{\nu}+2^{\nu-1}-1}\}\subset \Omega_{\tau^0}}
			K_{\omega^{j_1} , \omega^{j_2} , ..., \omega^{2^{\nu}+2^{\nu-1}-1}}\sigma_{\omega^{2^{\nu}+2^{\nu-1}-1}}) \cdot \\
			w_{ (\sigma_{\omega^1}=+1,\sigma_{\omega^{2}}=+1,...,,\sigma_{\omega^{2^{\nu}+2^{\nu-1}-2}}=+1,\sigma_{\omega^{2^{\nu}+2^{\nu-1}-1}})})  +\\
			\ln(	w_{ (\sigma_{\omega^0}=+1,\sigma_{\omega^{1}}=+1,...,\sigma_{\omega^{2^{\nu}+2^{\nu-1}-2}}=+1)  })- 	
			\ln(w_{ (\sigma_{\omega^0},\sigma_{\omega^{1}},\sigma_{\omega^{2}},...,	\sigma_{\omega^{2^{\nu}+2^{\nu-1}-2}})  }  ).
		\end{gathered} 	
	\end{equation} 
		
	By revising $\nu$ - dimensional models with the  Hamiltonian, containing nonzero only the interaction coefficients of the product of an even number of spins, we need to take free parameters in (\ref{A_ANNNI_nuD}),  contain ${ \omega^{2^{\nu}+2^{\nu-1}-1}}$, and corresponding to the interaction of an odd number of spins equal to zero:
	$	K_{\omega^{j_1} , \omega^{j_2} , ..., \omega^{j_{2 p}} , \omega^{2^{\nu}+2^{\nu-1}-1}}=0 $ .  The eigenvector $\overrightarrow{w}$ should look like (\ref{w_ANNNI_nu}) and satisfy the central symmetry condition (\ref{w_even})  . Then from the conclusions of item (\ref{even}) in the formula  (\ref{K_ANNNI_nuD}) the interaction coefficients of the product of an odd number of spins are equal to zero: 	$	K_{\omega^{j_1} , \omega^{j_2} , ..., \omega^{j_{2 p-1}}}=0 $ for ${\{\omega^{j_1},\omega^{j_2},...,\omega^{j_{2 p -1}}\}\subset \Omega'_{\tau^0}}. $ 
	We can simplify the formula (\ref{K_even}) for this case:

	\begin{equation}\label{K_even_ANNNI}
		\begin{gathered}			
			K_{\omega^{j_1} , \omega^{j_2} , ..., \omega^{j_{2 p}}}=\\
			-(\sum_{\{\sigma_{\omega^0},\sigma_{\omega^1},...,\sigma_{\omega^{2^{\nu}+2^{\nu-1}-3}}\}}
			\sigma_{\omega^{j_1}}\sigma_{\omega^{j_2}}...\sigma_{\omega^{j_{2 p}}} \cdot	\\
			A_{\sigma_{\omega^0} ,\sigma_{\omega^1},...,\sigma_{\omega^{2^{\nu}+2^{\nu-1}-3}},
				\sigma_{\omega^{2^{\nu}+2^{\nu-1}-2}}=+1 })/2^{2^{\nu}+2^{\nu-1}-2} \;\;,\\
			\{\sigma_{\omega^0} ,\sigma_{\omega^1},...,\sigma_{\omega^{2^{\nu}+2^{\nu-1}-2}} \} 	\neq
			(+1,+1,...,+1 )	,
			{\{\omega^{j_1},\omega^{j_2},...,\omega^{j_{2 p}}\}\subset \Omega'_{\tau^0}},\\ 	
			K_{\omega^{j_1} , \omega^{j_2} , ..., \omega^{j_{2 p-1}}}=	0,\\
			{\{\omega^{j_1},\omega^{j_2},...,\omega^{j_{2 p}}\}\subset \Omega'_{\tau^0}},	  
			{\{\omega^{j_1},\omega^{j_2},...,\omega^{j_{2 p -1}}\}\subset \Omega'_{\tau^0}},	 				
		\end{gathered} 	
	\end{equation} 
	
	where   $A_{\sigma_{\omega^0} ,\sigma_{\omega^1},...,\sigma_{\omega^{2^{\nu}+2^{\nu-1}-3}},
		\sigma_{\omega^{2^{\nu}+2^{\nu-1}-2}}=+1 } $ we find by the formulas (\ref{A_right}) with a centrally symmetric vector $	\overrightarrow{ w}$ (\ref{w_even}).

		%%%%%%%%%%%%%%%%%%%%%%%%%%%%%%%%%%%%%%%%%%%%%%%%%%%%%%%%%%%%%%%%% 

	\end{document}